\documentclass[fleqn,usenatbib,useAMS]{mnras}

\usepackage{newtxtext,newtxmath}

\usepackage[T1]{fontenc}
\usepackage{ae,aecompl}
\usepackage{color}
\usepackage{xspace}
\usepackage{rotating}
\usepackage{lscape}
\usepackage{float}

\usepackage{graphicx}	
\usepackage{amsmath}	
\usepackage{amssymb}	
\usepackage{multirow}
\usepackage{array,booktabs}
\usepackage[flushleft]{threeparttable}
\usepackage{siunitx}
\usepackage[english]{babel}
\usepackage[utf8x]{inputenc}
\usepackage{ulem}

\newcommand{\revi}[1]{{#1}}
\newcommand{\ignore}[1]{} 

\renewcommand{\star}{*}

\newcommand{\ie}{{\it i.e.}\xspace}
\newcommand{\eg}{{\it e.g.}}

\newcommand{\md}{\mathrm{d}} 
\newcommand*\dif{\mathop{}\!\mathrm{d}}
\newcommand{\pcc}{~{\rm cm}^{-3} }	

\newcommand{\HI}{\mbox{H\,{\sc i}}\xspace}
\newcommand{\HII}{\mbox{H\,{\sc ii}}\xspace}
\newcommand{\msun}{{\rm M}_{\odot}}
\def\tff{t_{\rm ff}}
\def\fesc{$\langle f_{\rm esc}\rangle$\xspace}
\def\hide#1{}

\begingroup
\catcode`\_=\active
\gdef_#1{\ensuremath{\sb{\mathrm{#1}}}}
\endgroup
\mathcode`\_=\string"8000
\catcode`\_=12

\title[Star Clusters Across Cosmic Time]
{Simulating Star Clusters Across Cosmic Time: I. Initial Mass Function, Star Formation Rates and Efficiencies}

\author[C.-C. He, M. Ricotti, S. Geen]{
Chong-Chong He,$^{1}$\thanks{E-mail: chongchong@astro.umd.edu}
Massimo Ricotti,$^{1}$
and Sam Geen$^{2}$\\
$^{1}$Department of Astronomy, University of Maryland, College Park, MD, 20742, US\\
$^{2}$Universit\"at Heidelberg, Zentrum f\"ur Astronomie, Institut f\"ur Theoretische Astrophysik, Albert-Ueberle-Str. 2, 69120 Heidelberg, Germany
}

\date{Accepted XXX. Received YYY; in original form ZZZ}

\pubyear{2019}

\begin{document}
\label{firstpage}
\pagerange{\pageref{firstpage}--\pageref{lastpage}}
\maketitle

\begin{abstract}
We present radiation-magneto-hydrodynamic simulations of star formation in self-gravitating, turbulent molecular clouds, modeling the formation of individual massive stars, including their UV radiation feedback. The set of simulations have cloud masses between $m_{\rm gas}=10^3$~M$_\odot$ to $3 \times 10^5$~M$_\odot$ and gas densities typical of clouds in the local universe ($\overline n_{\rm gas} \sim 1.8\times 10^2$~cm$^{-3}$) and 10$\times$ and 100$\times$ denser, expected to exist in high-redshift galaxies.
The main results are: {\it i}) The observed Salpeter power-law slope and normalisation of the stellar initial mass function at the high-mass end can be reproduced if we assume that each star-forming gas clump (sink particle) fragments into stars producing on average a maximum stellar mass about $40\%$ of the mass of the sink particle, while the remaining $60\%$ is distributed into smaller mass stars. Assuming that the sinks fragment according to a power-law mass function flatter than Salpeter, with log-slope $0.8$, satisfy this empirical prescription. {\it ii}) The star formation law that best describes our set of simulation is $d\rho_*/dt \propto \rho_{gas}^{1.5}$ if $\overline n_{gas}<n_{cri}\approx 10^3$~cm$^{-3}$, and $d\rho_*/dt \propto \rho_{\rm gas}^{2.5}$ otherwise. The duration of the star formation episode is roughly $6$ cloud's sound crossing times (with $c_s=10$~km/s). {\it iii}) The total star formation efficiency in the cloud is $f_*=2\% (m_{\rm gas}/10^4~M_\odot)^{0.4}(1+\overline n_{\rm gas}/n_{\rm cri})^{0.91}$, for gas at solar metallicity, while for metallicity $Z<0.1$~Z$_\odot$, based on our limited sample, $f_*$ is reduced by a factor of $\sim 5$. {\it iv)} The most compact and massive clouds appear to form globular cluster progenitors, in the sense that star clusters remain gravitationally bound after the gas has been expelled.
\end{abstract}





\section{Introduction}
\label{sec:intro}

Stars formation in galaxies is a complex and only partially understood astrophysical phenomenon. It is difficult to formulate a general theory in part because of the wide range of scales and of physical processes involved.
From an observational point of view, quantifying star formation efficiency (SFE) in nearby molecular clouds has been the focus of much recent research \citep[e.g.,][]{lada2010star,heiderman2010star,Gutermuth:2011}.
A power-law relationship between the gas surface density of galaxies and their star formation rate (SFR) was first proposed by \cite{Schmidt:1959} and later tested by large, multi-galaxy data \citep{Kennicutt:1998}. This relationship has been widely used in cosmological simulations of galaxy formation. However, on sub-galactic scales the dispersion of star formation rates for a given gas surface density of \HI is large, and other parameters such as gas metallicity \citep{Bolatto:2011,Krumholz:2013} and stellar surface density \citep{Leroy:2008} appear to become important.
As the resolution of surveys improved, numerous studies have shown that star formation on kpc scales is more strongly correlated with H$_2$ surface density \hbox{\citep[e.g.][]{Krumholz:2014}} rather than atomic gas. 
Therefore, modern cosmological simulations of galaxy formation aim at reproducing the molecular phase of the interstellar medium (ISM) and adopt an empirical sub-grid recipe for star formation within partially resolved molecular clouds of the form $\dot \rho_* \propto \rho_{H_2}^n$, where typically $n=1$ or $1.5$.
The molecular phase of the ISM is treated in simulations using different prescriptions: \hbox{\cite{Robertson:2008}} pre-computed a grid of models from a photo-chemistry code,
\cite{Gnedin:2009} directly solved the formation and dissociation equations for H$_2$, but with and increased formation rate to model unresolved clumping, and \hbox{\cite{Kuhlen:2012}} used an analytic model to estimate the equilibrium H$_2$ abundance. The sub-grid recipe (with grid maximum resolution typically between few parsecs to few kpc) is calibrated to reproduce observational data in galaxies at $z=0$.

However, the conditions in the ISM of high-redshift galaxies are likely different to those found in the present day.
\revi{\cite{Krumholz:2012a} argue that the SFR is in fact correlated to the local free-fall time set by the gas density, not to the column density.}
Simulations show that densities and pressures of star-forming regions in high-redshift galaxies are much higher than in today's ISM \citep[e.g.,][]{Ricotti:2002,Wise:2014,Ricotti:2016}. Using Adaptive Mesh Refinement (AMR) simulations of the first stars and galaxies with parsec-resolution, \cite{Ricotti:2016a} found that  compact molecular clouds in primordial galaxies can either form gravitationally bound star clusters that resemble the progenitors of today's globular clusters (GC)\footnote{The compact bound stellar objects found in the simulations are actually not only globular clusters progenitors, but also ultra-compact dwarfs and dwarf-globular transition objects, depending on whether the stellar cluster forms at the centre of the halo, in the disk's spiral arms, or even in satellite minihalos.}, or the clusters may disperse and fill up a large fraction of the dark matter halo of primordial dwarf galaxies. In this second case the stars would appear as spheroids 20-200~pc in radius, dark matter dominated and with very low surface brightness. These objects would be identified today as ``ultra-faint'' dwarf galaxies observed in the Local Group \citep[\eg,][]{Willmanetal05ApJ, Zuckeretal06a, Zuckeretal06b, Belokurovetal07, Walshetal07, Majewskietal07, Martinetal09}. Star formation in compact star clusters appears to be especially important, even perhaps the dominant mode of star formation at high-redshift. Thus, in order to make progress in understanding the formation of the first dwarf galaxies and the sources of reionisation, it is important to focus on understanding the small-scale physics of this process, which is poorly resolved in cosmological simulations.

Most numerical work on star formation in molecular clouds focuses on star formation in the local universe, aiming at explaining observed young star forming regions.
In this paper we analyse the results of a large grid of simulations of realistic molecular clouds with initial conditions chosen to reproduce not only local molecular clouds but also clouds that form in higher density and pressure environments, typical of star formation in high redshift galaxies. We vary the masses of the clouds, their compactness (central density), and in few cases explore the effect of changing the gas metallicity and therefore the gas cooling function.  

The motivation for this paper is twofold. The first goal is to deepen our understanding of the physics of star formation in high-pressure environments to justify and inform the sub-grid star formation recipe used in cosmological simulations. A closely related important question in Near Field Cosmology is: how does the formation of self-gravitating bound star-clusters relate to the star formation efficiency, compactness, mass and gas metallicity of molecular clouds found in cosmological simulations? We will only touch on this questions in the present paper, but more detailed work will be presented in a followup paper.

The second goal is to estimate the escape fraction of \HI ionising radiation from molecular clouds as a function of cloud compactness and mass. This is the first necessary step for a realistic estimate of the escape fraction from galaxies. \cite{Ricotti:2002} have shown that, if a non-negligible fraction of today's GCs formed at $z>6$ with \fesc$\sim 1$, their progenitors would be a dominant source of ionising radiation during reionisation. \cite{Katz:2014} presented arguments in support of significant fraction of today's old GCs forming before the epoch of reionisation. However, although it is naively expected, it has not been shown with numerical simulations that \fesc from GC progenitors forming in compact molecular clouds is higher than \fesc in more diffuse clouds. The answer to this question and the contribution of compact star clusters to reionisation will be presented in a separate companion paper.

This paper is organised as follows. In Section~\ref{sec:review} we present a brief review of the current status of numerical simulations of star cluster formation. In Section~\ref{sec:2} we provide an overview of our numerical methods, including  details on the initial conditions of our simulations and the recipes for formation of sink particles and feedback. In Section~\ref{sec:analyse} we present some results from the analysis of our large set of simulations with emphasis on the stellar initial mass function (IMF) and the star formation rate (SFR) and efficiency (SFE). A summary and conclusions are presented in Section~\ref{sec:summary}.

\subsection{The IMF and SFE of Molecular Clouds}
\label{sec:review}

Simulations of molecular cloud dynamics are valuable tools in understanding the conditions in the ISM. Typically these simulations adopt idealised initial conditions similar to those in observed clouds: a gas cloud $\sim 1-10$~pc in size supported against gravity by a turbulent velocity field such that the initial virial ratio, {\it i.e.} the ratio of the kinetic energy to the potential energy of the cloud, is $\lesssim 0.5$. One model involves injecting turbulence into a volume of gas in the initial conditions and allowing it to decay over time. This can be done by either using smoothed particle hydrodynamics \citep[SPH, e.g.][]{Klessen:2001,Bonnell:2006} or grid-based methods \citep[e.g.,][]{Gammie:2003}. Another model involves adding turbulence continuously over time, simulating the effect of momentum injection from outside flows or energy from massive stars inside the cloud \citep{Vazquez-Semadeni:1997,Ballesteros-Paredes:2006,Padoan:2007}. Many of these models adopted an isothermal equation of state, while others have included self-consistent cooling and heating functions \citep[e.g.][]{Koyama2004,Audit2005} and molecular chemistry \citep[e.g.][]{Glover2010}.

The fragmentation of molecular clouds into stars is a long-standing problem. Observational studies \citep{Salpeter:1955,Kroupa:2002,Chabrier:2005} have found that the masses of stars follow a ``Initial Mass Function'' (IMF) with a power law $\dif N / \dif \log M \propto M^{-\Gamma}$ at the high-mass end (\cite{Salpeter:1955} calculate $\Gamma \approx 1.35$). Various theoretical models have been constructed to explain this \citep{Padoan2002,MacLow2004,Hennebelle2008,Hopkins:2012a} based on gravoturbulent fragmentation of the host cloud. Radiative stellar feedback has been invoked to explain the precise shape of the IMF, using both simulations \citep[e.g.,][]{Bate:2009} and analytic models \citep[e.g.,][]{Guszejnov2016}.
Early pioneering simulations of cluster formation approached the problem of producing a well-defined IMF
\citep[e.g.,][]{Bate:2003,Bate:2005,Klessen:2008,Offner:2008}, but were often limited in terms of statistics or resolution. More recent work, with increasing computing power, provided more reliable statistics and IMF distributions \citep[e.g.,][]{Bonnell:2003,Bate:2009,Bonnell:2011,Girichidis:2011,Krumholz:2011,Bate:2012,Ballesteros-Paredes:2015}.

Simulations attempting to capture the stellar IMF require a high dynamic range to resolve both brown dwarfs and OB stars. Most recently, \cite{Bate:2019} resolve in detail the mass spectrum of brown dwarfs while only producing stars of up to 3 $\msun$, finding that low metallicities do not produce observable differences in the stellar IMF, while increasing fragmentation.  \cite{Gavagnin:2017} have lower mass resolution but capture more massive stars that emit significant quantities of ionising radiation, arguing that this alters the high mass end of the IMF. In the absence of radiation and cooling, \cite{Lee2018a} and \cite{Lee2018} study the early formation of protostellar Larson cores, and find that the choice of equation of state (eos) has a strong influence on the peak of the IMF. In general, these works are relatively successful at reproducing not only the IMF but also stellar multiplicity and separation.

Previous authors have included ideal MHD in their simulations \citep{Myers2013,Krumholz2016,Cunningham2018}. However, since these authors only form stars up to $\sim$20 $\msun$, they neglect ionising radiation. Non-ideal MHD effects, while challenging to include in resolving the IMF for reasons of computational cost, appear to affect the dynamics of protostar formation on small scales \citep{Masson2016,Vaytet2018}. The physics that shapes the IMF is complex, and a full treatment that covers non-ideal MHD, both low and high energy radiation, chemistry and the full mass range of stars remains difficult with modern computational resources.

As well as the stellar IMF, an important consideration is how many stars are formed out of a given mass of gas, or the Star Formation Efficiency (SFE). The efficiency of conversion of gas into stars is typically much lower than 100\% since energetic processes from massive stars are able to disperse the cloud in which a star cluster forms before all of the gas collapses into protostars. These processes are widely termed ``feedback'' \citep[see review by][]{Dale2015}. Recent work favours ionising radiation as the main driver of molecular cloud dispersal \citep{Dale2005,Gritschneder2009,Peters2010,Walch2012,Dale2012}, as opposed to other effects such as stellar winds \citep{Dale:2014}, although \cite{Howard:2016} find that UV photoionisation has little effect on the initial evolution of the SFE.

The relationship between gas properties and the SFE is a matter of ongoing study. \cite{lada2010star} and \cite{heiderman2010star} argue for a constant ratio between gas above a certain surface or column density, although this is still subject to discussion \citep{Gutermuth:2011, Hony2015}. There is no clear theoretical link between the SFE and projected column density, although \cite{Clark2014} argue that there may be a link between the observed column density and the local density around the star. \cite{Geen:2017} reproduce the SFE observed by \cite{lada2010star}, although they find that the result is likely to be dependent on the average density of the neutral gas in the cloud. These simulations produce similar results to the simulations of \cite{Colin2013}. \cite{Geen2018} finds that SFE can change by up to a factor of 4 by varying the initial velocity field of the cloud and the stellar IMF, although relationships can be found between the early cloud state and the final SFE. Semi-analytic models by \cite{Vazquez-Semadeni2018} also find considerable scatter in the SFE.


\section{Numerical Simulations and Methods}
\label{sec:2}

We conduct our numerical simulations using the AMR radiative magneto-hydrodynamical code \textsc{ramses} \citep{Teyssier:2002, Bleuler:2014}. 
Radiative transfer is implemented using a first-order moment method \revi{with M1 closure} described in \cite{Rosdahl:2013}.
\revi{
\cite{Kim:2017} demonstrates that M1 closure method is inaccurate near sources only in regions where the flux is about an order of magnitude smaller than the mean value (due to shielding), while it agrees with adaptive ray-tracing methods \citep[\eg,][]{Wise:2014,Hartley:2016} both at larger distances from individual sources and on global scales. M1 closure, however, is significantly more computationally efficient than ray-tracing methods.
}
The ionising photons interact with neutral gas and we track the ionisation state and cooling/heating processed of hydrogen and helium \citep[see][for details]{Geen:2017}. Our simulations include magnetic fields in the initial conditions, but we do not include the chemistry of molecular species (\ie, formation/dissociation).
3-D `zoom-in' simulations of the chemical evolution of molecular clouds suggest that, for gas at solar metallicity, the cloud is almost fully molecular with H2 fractions around 0.9 in the later stages of transition to dense molecular phase \citep{Seifried:2017}. 

We simulate a set of isolated and turbulent molecular clouds that collapse due to their own gravity. We explore a grid of simulations varying the initial gas mass and compactness (\ie, the core density) of the clouds.  In our simulations, dense proto-stellar cores collapsing below the resolution limit of the simulations produce sink particles. These sinks may represent single stars or multiple stars or even clusters of stars if the resolution is not sufficiently high. However, in all our simulations we aim at reproducing a realistic high-mass end of the stellar IMF and therefore realistic feedback from individual massive stars. To accomplish this goal, sink particles emit hydrogen and helium ionising photons according to their mass as described in \S~\ref{sec:uv}. The gas is ionised and heated by massive stars, producing over-pressurised bubbles that blow out the gas they encounter. In our simulations low mass stars and proto-stellar cores do not produce any feedback.
In this work we do not include mechanical feedback from supernova (SN) explosions and from stellar winds and we also neglect the effect of radiation pressure from infrared radiation. However, with the exception of the two most massive clouds in the set of simulations representing today's molecular clouds (the lowest density set), all the simulations stop forming stars before the explosion of the first SN. Therefore, neglecting SN feedback is  well justified in these cases.
\revi{
We find that in all simulations star formation has ceased before $\sim 5 - 6 ~\tff$, which is the typical time it takes for feedback to act. We stop simulations no earlier than this point. A few simulations are continued beyond this time. This does not have an affect on the IMF since the mass function of sink particles does not change after star formation has ceased.
}

For the simulations in the set in which SN explosions should occur while star formation is ongoing, in order to compensate for this missing feedback, we do not shut down UV radiation feedback from massive stars after the time the star should have exploded as SN. In the following sections we provide some more details on the simulations set up.

\subsection{Initial Conditions}
\label{sec:ini}

We run a grid of 14 simulations of clouds with a range of central densities and initial gas masses. We also run some additional simulations varying the initial gas metallicity and therefore the gas cooling function. The magnetic field strength in the initial conditions is set such that $v_a = 0.2~\sigma_{3D}$, where $v_a$ is Alfven wave velocity and $\sigma_{3D}$ is the turbulence velocity dispersion. \revi{This $\nu_a$ is $\sim 2$ times smaller than that measured in a group of molecular clouds by \cite{Crutcher:2012} who finds $\nu_a \approx 0.5 \sigma_{3D}$.}

The clouds have initially a spherically symmetric structure with density profile of a non-singular isothermal sphere with core density $n_{\rm c}$. The cloud extends out to $r_{\rm gas}=3r_{\rm c}$, where $r_{\rm c}$ is the core radius. Beyond $r_{\rm gas}$ the cloud is embedded in a uniform density envelope that extends to $6r_{\rm c}$ with a density $0.01~n_{\rm c}$. Outside of the envelope the number density is constant at $1$~cm$^{-3}$. The box length $L_{box}$ is set to $48 r_{\rm c}$ in each simulation. The initial value of the (isothermal) sound speed of the cloud is set to $c_{\rm s} = 0.24$~km/s, while the envelope and background densities are in pressure equilibrium.

The initial density profile is perturbed with a turbulent velocity field, analogously to the set up used in \cite{Geen:2017}. The initial turbulence of the clouds follows a Kolmogorov power spectrum with random phases and has an amplitude such that the cloud is approximately in virial equilibrium. All simulations have the same set of random phases. The initial cloud virial ratio
\begin{equation}
	\alpha_{\rm vir} = \frac{5\sigma_{\rm 3D}^2 R}{3GM} \approx 0.4,
\end{equation}
is kept constant in all the simulations. \revi{Therefore the ratio $t_{ff}/t_{turb}$, where $t_{turb}\equiv R/\sigma_{3D}$ is kept constant in all the simulations. However, the sound crossing time $t_{cr}\equiv R/c_s$, where troughout this paper we assume $c_s=10$~km/s, is not constant.} The virial parameter, $\alpha_{\rm vir}$, is small enough to ensure collapse and fragmentation, but sufficiently large to prevent a rapid radial collapse of the cloud. Before allowing any star formation in the cloud we evolve these idealised initial conditions for $\sim 3t_{\rm ff}$, so that the turbulent velocities develop into density perturbations and the initial conditions relax into a quasi-equilibrium turbulent medium. If we do not allow the initial conditions to relax before forming stars, the stars form mostly near the centre of the cloud during the transient relaxation phase.

\begin{table*}
\caption{
  Initial conditions of our 16 simulations.
}
\label{tab:1}
\begin{threeparttable}
\centering

\begin{tabular*}{0.9\textwidth}{@{\extracolsep{\fill}} c|r*{6}{c}}
&
$m_{\rm gas} (\msun)$ \tnote{d} &
$1.0 \times 10^3$ & 
$3.2 \times 10^3$ & 
$1.0 \times 10^4$ & 
$3.2 \times 10^{4} $ &
$1.0 \times 10^{5} $ & 
$3.2 \times 10^{5} $ \\
\hline

&
Cloud Name \tnote{e} &
&
XS-F &
S-F & 
M-F &
L-F & 
XL-F
\\
$\overline{n}_{\rm gas}$\tnote{a} $=1.8 \times 10^{2}\,$cm$^{-3}$ &
$r_{\rm gas}$ (pc) \tnote{f} &
&
5.0 &
7.3 &
11 &
16 &
23
\\
&
$\Sigma$ ($\msun$ pc$^{-2}$) \tnote{x} &
&
41 &
61 &
89 &
131 &
193
\\
&
$v_{esc}$ (km/s) \tnote{x} &
&
2.3 &
3.4 &
5.1 &
7.4 &
11
\\
&
$\Delta x_{\rm min}$ (AU) \tnote{g} &
&
500 &
730 &
1100 &
1600 &
2300
\\
$t_{\rm ff}$\tnote{b}\;$=4.4\,$Myr & 
$n_{\rm sink}$ (cm$^{-3}$) \tnote{h} &
&
$1.2\times 10^{7}$ &
$5.6\times 10^{6}$ &
$2.6\times 10^{6}$ &
$1.2\times 10^{6}$ &
$5.6\times 10^{5}$
\\
&
$M_{\rm J}~(M_\odot$) \tnote{i} &
&
0.3 &
0.4 &
0.6 &
0.9 &
1.3
\\
$l_{\rm max}$\tnote{c} $=15$ &
$\mathcal{M}$ \tnote{j} &
&
4.6 &
6.8 &
10 &
15 &
22
\\
&
$t_{\rm cr}$ (Myr) \tnote{k} &
&
0.5 &
0.7 &
1.1 &
1.5 &
2.3
\\
&
$Z/Z_\odot$ \tnote{l} &
&
1 &
1 &
1 &
1 &
1
\\
\hline

&
Cloud Name &
&
XS-C &
S-C & 
M-C &
L-C, L-C-lm, L-C-xlm \tnote{m} & 
\\
$\overline{n}_{\rm gas} = 1.8 \times 10^{3}\,$cm$^{-3}$ &
$r_{\rm gas}$ &
&
2.3 &
3.4 &
5.0 &
7.3 &
\\
&
$\Sigma$ &
&
193 &
283 &
415 &
609 &
\\
&
$v_{esc}$ &
&
3.4 &
5.1 &
7.4 &
11 &
\\
&
$\Delta x_{\rm min}$ &
&
460 &
680 &
1000 &
1500 &
\\
$t_{\rm ff}=1.4\,$Myr  &
$n_{\rm sink}$ &
&
$1.4\times10^7$ &
$6.5\times10^6$ &
$3.0\times10^6$ &
$1.4\times10^6$ &
\\
&
$M_{\rm J}$  &
&
0.3 &
0.4 &
0.6 &
0.8 &
\\
$l_{\rm max}=14$ &
$\mathcal{M}$  &
&
6.8 &
10 &
15 &
22 &
\\
&
$t_{\rm cr}$ &
&
0.23 &
0.33 &
0.5 &
0.7 &
\\
&
$Z$ &
&
1 &
1 &
1 &
1, $1/10$, $1/40$ &
\\
\hline

&
Cloud Name &
XXS-VC &
XS-VC &
S-VC & 
M-VC &
L-VC & 
\\
$\overline{n}_{\rm gas} = 1.8 \times 10^{4}\,$cm$^{-3}$ &
$r_{\rm gas}$ &
0.7 &
1.1 &
1.6 &
2.3 &
3.4 &
\\
&
$\Sigma$ &
609 &
894 &
1312 &
1925 &
2827 &
\\
&
$v_{esc}$ &
3.4 &
5.1 &
7.4 &
11 &
16 &
\\
&
$\Delta x_{\rm min}$ &
150 &
220 &
320 &
460 &
680 &
\\
$t_{\rm ff}=0.44\,$Myr  &
$n_{\rm sink}$  &
$1.4\times10^8$ &
$6.5\times10^7$ &
$3.0\times10^7$ &
$1.4\times10^7$ &
$6.5\times10^6$ &
\\
&
$M_{\rm J}$  &
0.08 &
0.12 &
0.17 &
0.26 &
0.38 &
\\
$l_{\rm max}$ $=14$ &
$\mathcal{M}$  &
7 &
10 &
15 &
22 &
32 &
\\
&
$t_{\rm cr}$ &
0.07 &
0.10 &
0.15 &
0.23 &
0.33 &
\\
&
$Z$ &
1 &
1 &
1 &
1 &
1 &
\\
\hline

\end{tabular*}

\begin{tablenotes}
\item[] 
  (a) Mean number density of the cloud, excluding the envelope. The core density is $\sim 5$ times higher.
  (b) The global free-fall time of the cloud ($t_{ff} \equiv 3\sqrt{\frac{3\pi}{32G\rho_{\rm c}}} \approx 1.3 \sqrt{\frac{3\pi}{32G\overline{\rho}}}$.
  (c) Maximum level of refinement.
  (d) Initial cloud mass, excluding the envelope.
  (e) The name of each cloud used throughout the paper. See Sec.~\ref{sec:ini} on how they are defined.
  (f) Initial cloud radius, excluding the envelope.
  (g) Maximum spatial resolution. 
  (h) Density threshold for sink formation. 
  (i) Jeans mass at the sink density threshold.
  (j) Turbulence Mach number.
  (k) Sound crossing time $r_{\rm gas}/c_s$ for $c_s=10$~km/s.
  (l) Metallicity of the gas used in the cooling function, $Z$~=~[Fe/H].
  (m) This setup has 2 extra simulations with lower metallicities
  besides one with same metallicity as all other ones. See Sec.~\ref{sec:met}.
  (x) The mean surface density in a square of the size of the cloud radius.
  (x) Escape velocity at the cloud radius of the initial cloud.
\end{tablenotes}
\end{threeparttable}
\end{table*}

\begin{figure}
    \centering
    \includegraphics[width=\columnwidth]{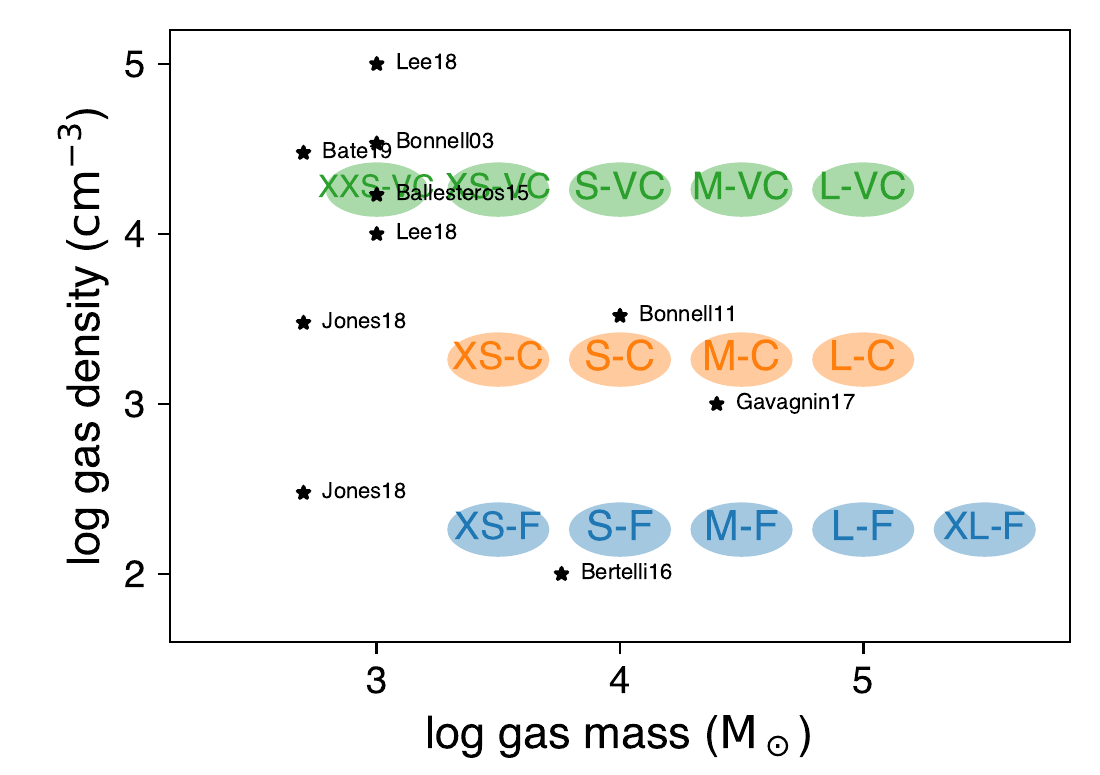}
    \caption{Simulation parameters in this work (colored ovals) compared to previous works (stars). The parameter space considered here is mass of the gas cloud (x-axis) versus mean particle number density of the cloud (y-axis).
    The labels showing the previous work found in the literature include: 
    \protect\cite{Bonnell:2003,Bonnell:2011,Ballesteros-Paredes:2015,Bertelli-Motta:2016,Gavagnin:2017,Jones:2018,Lee:2018,Bate:2019}.
    }
    \label{fig:comparison}
\end{figure}

A detailed list of the parameters in our simulations is shown in Table~\ref{tab:1}.
The clouds in Table~\ref{tab:1} are labelled with letters of two or three parts. The first part is either `XXS' (extra-extra-small), `XS' (extra-small), `S' (small), `M' (medium), `L' (large), or `XL' (extra-large), representing various initial gas masses of $\num{e3}$, $\num{3.16e3}$, $\num{e4}$, $\num{3.15e4}$, $\num{e5}$, and $\num{3.16e5}$~M$_\odot$, respectively.
The second part is either `F' (fiducial, which are the most similar to clouds in the solar neighbourhood), `C'(compact), or `VC' (very compact), in order of increasing initial mean gas density. 
The mean particle number density of the cloud, $\overline{n}_{gas} = \overline{\rho}_{gas}/(\mu m_p)$, where $\mu = 1.4$ is the mean molecular weight of the atomic gas, increases by a factor of ten between each set of simulations from $\num{1.8e2}$~cm$^{-3}$ to $\num{1.8e4}$~cm$^{-3}$. 
The `L-C' setup has two more simulations that have a third part in the name, `lm' and `xlm', representing `low-metallicity' and `extra-low-metallicity'.
A comparison of our setups with the literature is shown in Figure~\ref{fig:comparison}.

\subsection{Resolution and Sink Formation}\label{ssec:sink}
We use a Cartesian grid with an octree structure with cells that we subdivide into $2^3$ child cells as the simulation evolves (``adaptive refinement'').
Our starting refinement level is $ \ell_{\rm min} = 7 $ (corresponding to $\Delta x= L_{\rm box}/2^7)$ and maximum level of refinement is $ \ell_{\rm max} = 15 $ for runs with the lowest mean density and $ \ell_{\rm max} = 14 $ for all the other runs. The resolution is therefore $\Delta x_{\rm min}=L_{\rm box}/2^{15}$ for the "fiducial" runs, which corresponds to resolutions between $500$~AU and $2300$~AU. The "compact" clouds have resolution between $460$~AU and $1500$~AU and the "very compact" clouds between $150$~AU and $680$~AU.
 
In order to resolve the Jeans length with $N$ grid cells it is required that
\begin{equation}
  \label{eq:jeans}
  \lambda_{\rm J}=c_{\rm s}\sqrt{\frac{\pi}{G\rho}} > N_{sink} 
  \Delta x.
\end{equation}
From Equation~(\ref{eq:jeans}), the Jeans length is resolved with at least $N_{sink}$ grid points if $\rho<\rho_J$, where
\begin{equation}
	\label{eq:jeans2}
    \rho_{\rm J} = \frac{\pi c_{\rm s}^2}{G N^2 \Delta x^2}.
\end{equation}
In our simulations we enforce the refinement criterion that the Jeans length is resolved with at least $N_{\rm ref}=10$ cells. Hence, when the local density goes up and reaches a point where $\lambda_{\rm J}$ becomes smaller than $N_{\rm ref} \Delta x$, each cell is refined individually into eight new children cells. This refinement condition is always true, up to the maximum refinement level (when $\Delta x=\Delta x_{\rm min}$).
When the gas density exceeds $\rho_{\rm J}^{\rm max}=\rho_{\rm J}(\Delta x=\Delta x_{\rm min},N)$ at the maximum refinement level, we cannot continue to resolve the Jeans length with at least $N_{\rm ref}$ cells. We therefore create sink particles to trace material above these densities. We set $\rho_{\rm sink}=\rho_{\rm J}(\Delta x=\Delta x_{\rm min},N=N_{\rm sink})$ as critical density threshold to form sink particles.
Sink particles are created on the fly using a peak detection algorithm \citep[see][for details on sink particle formation in RAMSES]{Bleuler:2014}. We first detect density clumps above a density threshold $f_{\rm c} \rho_{\rm sink}$, with $f_{\rm c} = 0.1$. Then, the algorithm performs a peak density check, a collapsing check ($\nabla \cdot v =0$), and virial check before forming a sink particle.

In order to avoid numerical fragmentation it is usually suggested that $N_{\rm sink} \ge 4 $ \citep{Truelove:1997}. 
In our simulations we adopt $N_{\rm sink}=5$ for reasons detailed Appendix~\ref{sec:app1}. With an initial sound speed $ c_{\rm s} = 0.24$~km/s the Jeans mass at the sink density threshold is
\begin{equation}
\label{eq:jeansmass}
  M_{\rm J} = \frac{4\pi}{3} \rho_{{\rm J}} \left( \frac{\lambda_{{\rm J}}}{2} \right)^3 \ignore{= \frac{\pi^2}{6G}  c_s^2 N_{\rm sink} \Delta x_{\rm min} }= 0.55 M_\odot \left( \frac{\Delta x_{\rm min}}{1000\,{\rm AU}} \right),
\end{equation}
which results in $M_{\rm J} \sim 0.08 M_\odot $--$ 0.8 M_\odot $ for the compact and very compact clouds and $\sim 0.3 M_\odot $--$ 1.3 M_\odot $ for the fiducial clouds.

The sink particles are then treated like point masses and accrete gas based on the mechanism described as `threshold accretion' in \cite{Bleuler:2014}. The dynamics of the sink particles takes into account gravitational force from gas and stars and it is evolved using a leap-frog integration scheme. The effect of gas dynamical friction is not included.

\subsection{Feedback and Properties of UV source}
\label{sec:uv}
In our simulations, ionising UV photons are emitted from sink particles from the time they form to the end of the simulation.
Massive stars have lifetime of few Myrs, shorter than the duration 
of some of our simulations, and they may explode as SNe during the simulation. Since we are not implementing SNe feedback, we
keep the stars emitting radiation after their death to compensate for the lack of SNe in the attempt of avoiding underestimating feedback effects.
While SN explosions produce a significant amount of mechanical energy (typically $10^{51}$~egs), the energy associated with ionising radiation from massive stars integrated through their main-sequence lifetime is comparable (or larger for more massive stars) and this feedback starts acting earlier than SN feedback. For an O-star, more than half of the radiation is emitted in hydrogen ionising photons. Typically $\sim 10\%$ of a star's hydrogen is burned in the nuclear fusion process, with an energy efficiency of $\sim 0.7\%$. Thus the amount of energy radiated by a massive star during its life time is \ignore{$\sim 10^{-3} M_\star$, or $\sim 2 \times 10^{52}$~ergs for a 10~M$_\odot$ star.}
\revi{$\sim 2 \times 10^{-3} M_\star$, or $\sim 4 \times 10^{52}$~ergs for a 20~M$_\odot$ star.}

For each simulation we estimate the total hydrogen-ionising photon emission rate at a given time as
$S_{\rm cl}(m_{cl}) = 8.96\times10^{46}\,\textrm{s}^{-1}\,(m_{cl}/M_\odot)$ \citep[see][]{Geen:2017}, where $m_{cl}$ is the total mass of the sink particles.
This is calculated by Monte Carlo sampling a stellar population as described in \cite{Geen:2016} (See Sec.~\ref{sec:app2}).
The fraction of the total hydrogen ionising photon emission rate attributed to each sink particle is based on the following relation
\begin{equation}
	q(m_{\it i}) = V(0.3m_{\it i})\left(\frac{S_{\rm cl}(\Sigma_{\it i} m_{\it i})}{\Sigma_{\it i} V(0.3m_{\it i})}\right),
\end{equation}
where $V(m)$ is the hydrogen-ionising photon emission rate from a star with mass $m$, using the fits from \cite{Vacca:1996}.

The factor $0.3$ is an empirical factor to account for the scaling between the masses of sink particles and those of massive stars, necessary because we do not fully resolve the fragmentation of sink particles into proto-stars. See Section~\ref{sec:imf} for further discussion. 
The correction factor $X\equiv S_{\rm cl}(\Sigma_i m_i)/\Sigma_i V(0.3m_i)$ is very close to unity in most simulation in which we resolve massive stars and it is introduced only to prevent overproducing ionising radiation in case massive stars are poorly resolved. In all simulations we also impose $X \le 1$.

For the fiducial clouds we also include He$^0$ and He$^+$ ionising photons with total emission rates being
$S_{\rm He^0} = (1.178\times10^{46}\,\textrm{s}^{-1})(M_\star/M_\odot)$ and $S_{\rm He^+} = (2.422\times10^{43}\,\textrm{s}^{-1})(M_\star/M_\odot)$.
These rates are calculated using the same method as for hydrogen ionising photons described above using a Kroupa IMF \citep{Kroupa:2002}.
For the luminosity of individual stars we use \cite{Schaerer:2002} fitting for $Q$(He$^0$) and $Q$(He$^+$) with extrapolations above 150~$M_\odot$. This is contrasted by the model of \cite{Gavagnin:2017}, who assume blackbody spectra for each star. See Appendix~\ref{sec:app2} for details.

\revi{Various authors have concluded that UV photoionisation is typically the most important process in regulating star formation on a cloud scale. \cite{Haworth:2015} find that additional processes beyond hydrogen photoionisation have a correcting factor of 10\% at best. Radiation pressure mainly becomes important at very high surface densities, which principally affects smaller scales than the ones studied here - see \cite{Crocker:2018} for idealised conditions and \cite{Kim:2018} for simulations with self-consistent star formation feedback. \cite{Dale:2014} further find that winds have a minimal effect on the star formation efficiency of molecular clouds. We are thus justified in our choice to focus on UV photoionisation feedback in this work, but discuss cases where this may not be sufficient later in the paper.
}

\subsection{Cooling}
\label{sec:cooling}

We use the radiative cooling function described in \cite{Geen:2016}. The cooling in neutral gas is based on the prescription in \cite{Audit2005}, which includes cooling from carbon, oxygen and dust grains as well as the effect of the ambient UV background in the ISM. For collisionally ionized gas at temperatures $>10^4$~K we use \cite{Sutherland1993} cooling function. The out-of-equilibrium cooling of photoionised hydrogen and helium is treated as described in \cite{Rosdahl:2013}. Out-of-equilibrium cooling of photoionised metals is treated with a piecewise fit to the cooling curve given in \cite{Ferland2003}. We assume a uniform metallicity as listed in Table \ref{tab:1}. For most simulations, this is solar metallicity, though we perform some simulations at sub-solar metallicity. We do not implement out-of-equilibrium molecular chemistry.

\section{Results}\label{sec:analyse}

In this section we present and discuss the results of our simulations. In \S~\ref{sec:imf}, we study the mass function of cores (sink particles) and the IMF. In \S~\ref{sec:sfe} we focus on the star formation efficiency and in \S~\ref{sec:sflaw}, on the star formation rate. 
\begin{figure*}
    \centering
    \includegraphics[height=.88\textheight]
    {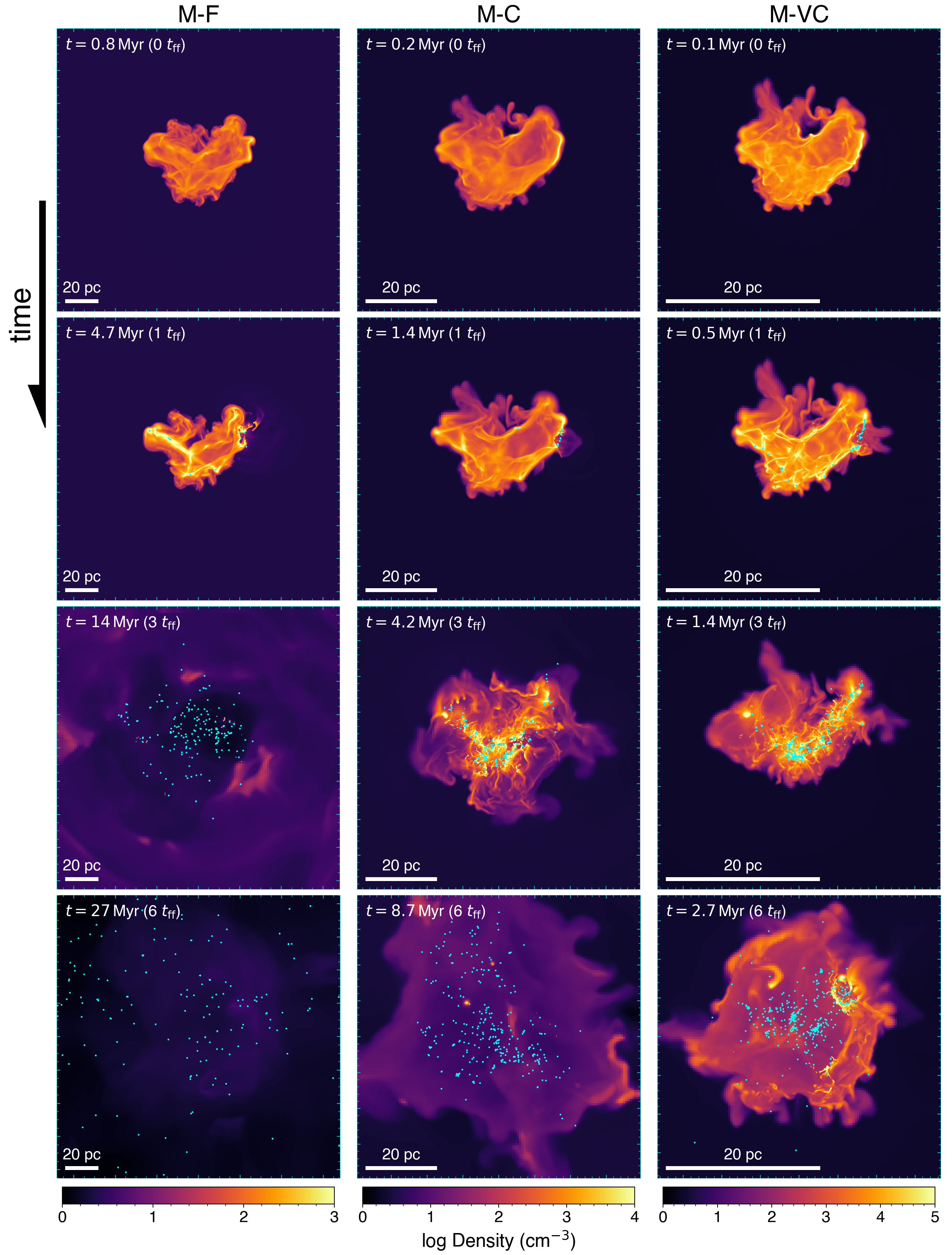}
    \caption{
    Line-of-sight projections of the (density-weighted) gas density for three simulations with cloud mass $3.2\times 10^4$~M$_\odot$. From left to right we show clouds with increasing mean density: $\overline n_{\rm gas} \sim 1.8\times 10^2$~cm$^{-3}$, representing our fiducial clouds in the local universe, $\overline n_{\rm gas} \sim 1.8\times 10^3$~cm$^{-3}$ and $\overline n_{\rm gas} \sim 1.8\times 10^4$~cm$^{-3}$, respectively. From top to bottom we show the time evolution of the clouds.
    Sink particles are displayed as cyan dots.
    The snapshots shown in the top row represent the initial conditions of the turbulent cloud: no stars have formed at this time because the highest density is below the threshold for star formation, but the cloud idealised initial conditions have been already evolved for $\sim 3~t_{ff}$ in order to develop a turbulent density field. In the bottom row snapshots, star formation has stopped and most of the gas has been expelled as a result of UV feedback from massive stars. 
    }
    \label{fig:snap1}
\end{figure*}
\begin{figure*}
    \centering
    \includegraphics[height=.88\textheight]
    {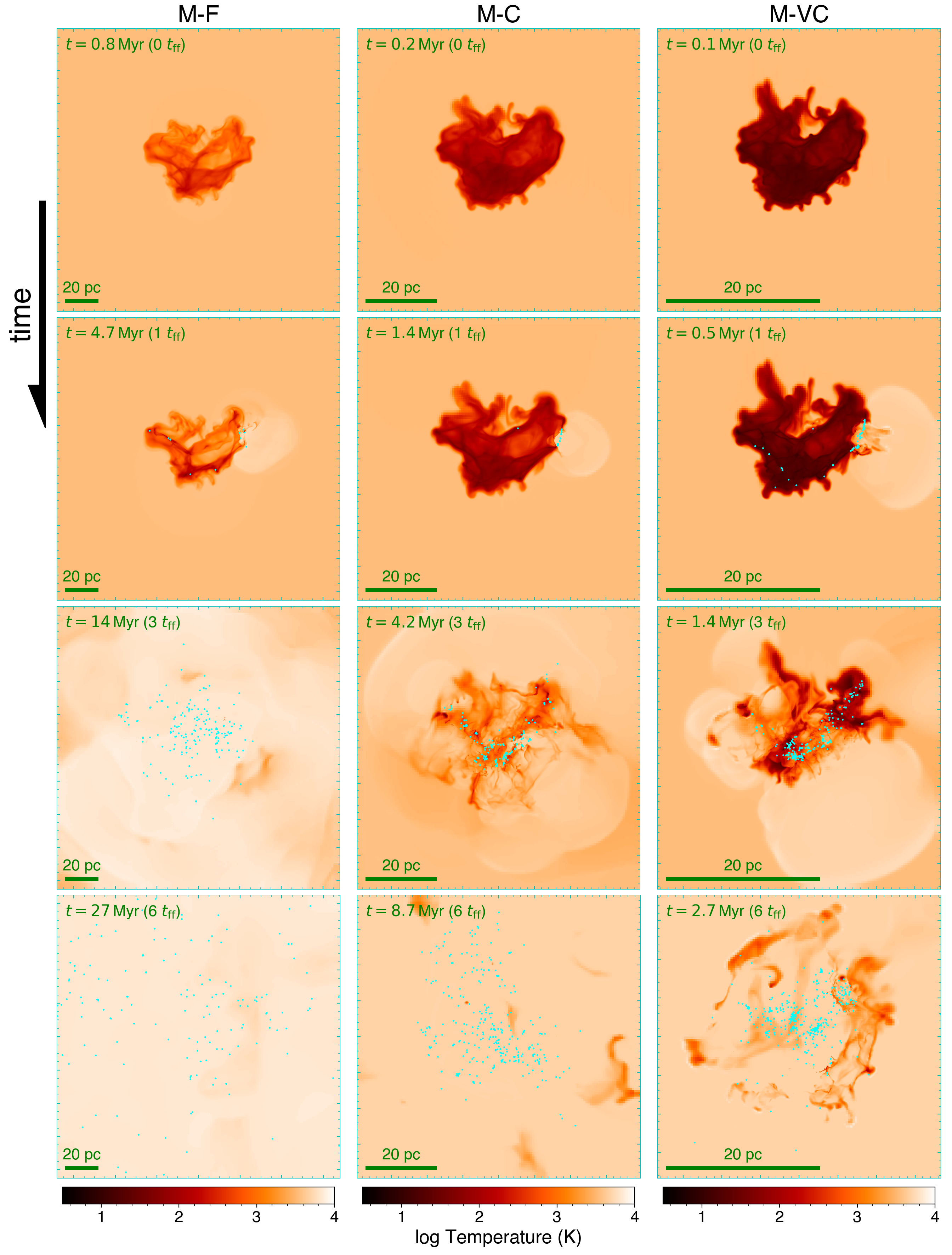}
    \caption{Same as Fig.~\ref{fig:snap1} but showing the density-weighted projection of the temperature.
    }
    \label{fig:snap2}
\end{figure*}
\begin{figure*}
    \centering
    \includegraphics[width=\textwidth]{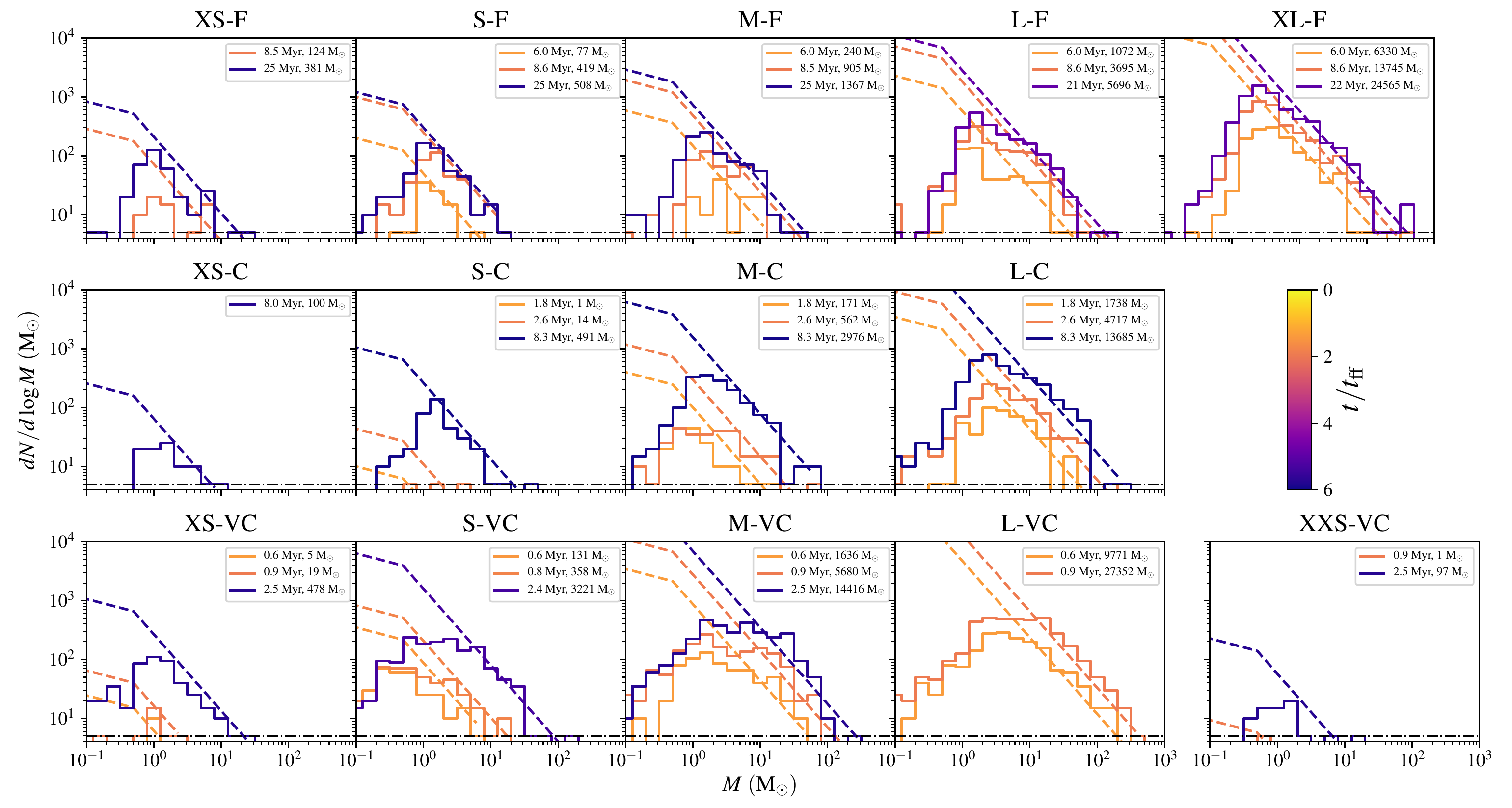}
    \caption{Temporal evolution of the IMF of stars, obtained multiplying by 0.4 the masses of the sink particles (see text).
  The dashed lines are analytic Kroupa IMF \revi{for systems} normalised to the total mass of the sink particles at the corresponding time. The time and total mass in sink particles are shown in the legend. We see good agreement between the shifted sink particles mass function and the analytic mass-normalised Kroupa IMF at the high-mass end, both in terms of the power-law slope and normalisation.
    }
    \label{fig:imf_evo}
\end{figure*}
A representative sample of snapshots from our simulation set is shown in Figures~\ref{fig:snap1} and \ref{fig:snap2}. 
These figures show the time evolution of the density-weighted projections of the gas density and temperature as well as the position of the sink particles along line of sight for three simulations with different mean clund densities (fiducial on the left column, compact, middle column, and very-compact on the right column), and a cloud mass of $3.2\times 10^4$~M$_\odot$.

We observe clearly that the star formation efficiency increases with increasing cloud density, and the stellar cluster that is formed at the end of the simulation remains more compact and self-gravitating for the densest cloud.
The effect of radiative feedback from massive stars is also clearly visible in the density and temperature projections. \HII regions break out of the dense filaments destroying them and reducing the overall mass in dense gas in which stars are formed.

\subsection{Stellar Initial Mass Function}\label{sec:imf}

\subsubsection{Cores Fragmentation and Initial Mass Function}\label{sec:CMF}

Maps in the continuum of cluster regions and larger areas in  star-forming systems allow to construct a core mass function (CMF), that is the mass function of high-density gas concentrations (starless cores) with mass sufficiently large to be identified given the resolution of the observations \citep[e.g.,][]{Motte:1998}.
Observations of a well-resolved CMF in the Pipe nebula show a striking similarity to the stellar IMF, but shifted to higher masses by a factor of a few, which suggests that the IMF is the direct product of the CMF with a roughly constant core-to-star conversion efficiency $\sim 30\%$ \citep[e.g.,][]{Matzner:2000, Alves:2007}

Previous works on star formation in molecular clouds which adopted sink particles (like in the present paper) have investigated the mapping between the masses of pre-stellar cores at the time they become self-gravitating and the final masses of the stars that form within them \citep[e.g.,][]{Padoan:2001,Smith:2009}. For instance, \cite{Smith:2009}, using SPH simulations, find that at early times the relationship between stellar masses and the parent cores can be reproduced within a modest statistical dispersion with the star being about one-third of the parent core mass. 

We find results in agreement with these previous studies.
Figure~\ref{fig:imf_evo} (and Figure~\ref{fig:imf_fit}) show the stellar mass function for our grid of simulations obtained assuming that the sink particles are about a factor of 2.5 more massive than the corresponding massive stars they produce. This means that the number of stars of a given mass ($>1$~M$_\odot$) is given by a Kroupa IMF for a star cluster with total mass equal to the total mass of the sinks.
In Figure~\ref{fig:imf_fit} the IMF is shown at the end of the simulation when star formation has stopped, while in Figure~\ref{fig:imf_evo} we also show the time evolution of the IMF, together with the Kroupa IMF for a cluster with total mass equal to the total mass in sink particles (dashed lines). \revi{This means that we are assuming nearly 100\% efficiency of star formation in the cores (\ie, the cores fragment into stars) or a lower efficiency but the gas expelled by feedback is later transformed into low-mass stars. However, we think that this second model is less physically motivated}.  We can see that the shifted sink mass function \revi{(SMF)} matches the Kroupa IMF at the high-mass end, both in terms of the slope and normalisation at any time during the formation of the star cluster.
The figure suggests that the birth of stars in a cluster follows the same random sampling of the universal mass function throughout the star-formation process.  In other words, it appears that there is not a bias toward formation of high mass-stars or low-mass stars during the early times when the cluster is in the formation process.

As discussed above the sink particles can be interpreted as pre-stellar cores, and each sink particle converts $\sim 40\%$ of its mass to a single massive star, with the rest of the mass ending up in low-mass stars, filling up the lower mass end of the IMF. The flattening or cut-off of the IMF at the low-mass end observed in our simulations is likely due to insufficient spatial and mass resolution to capture the formation of low mass cores or the fragmentation of more massive pre-stellar cores.
To further clarify, we note that the interpretation that only 40\% of the core mass is converted into a star and the remaining 60\% is returned to the gas phase, never to participate in star formation, would not produce the correct normalisation of the IMF. This is because in this scenario, although the stellar masses are $\sim 40\%$ of the core masses, the total stellar mass and the star formation efficiency would be reduced by a factor of $\sim 2.5$, lowering the expected number of massive stars below the value found in the simulations.

\revi{ 
So far we have assumed that all the gas in the cores fragments into stars with $\eta=100\%$ efficiency. In this case we find a conversion factor $\varepsilon = 0.4$ between the CMF and the IMF (such that the normalisation of the IMF agrees with the observed one). However, it is possible to match the observed IMF also in models in which $\eta<1$. Note that in this case the TSFE shown in all our plots should be re-scaled by a factor $\eta$. In models with $\eta<1$, the conversion factor $\varepsilon$ that matches the mass-normalised empirical IMF, is $\varepsilon = 0.4 \eta^{1/\Gamma}$. 
For $\eta = 1, 0.69$ and $0.4$, $\varepsilon = 0.4,0.3$ and $0.2$, respectively.
}

The results of this section justify our assumptions to model radiative feedback in Section~\ref{sec:uv} and it further implies that our simulations are self-consistently treating the formation of individual massive stars, their feedback effects, and can be used reliably to estimate of the escape fraction of hydrogen and helium ionising photons.

\subsubsection{\revi{Resolution Studies}}

\begin{figure*}
    \centering
    \includegraphics[width=\textwidth]{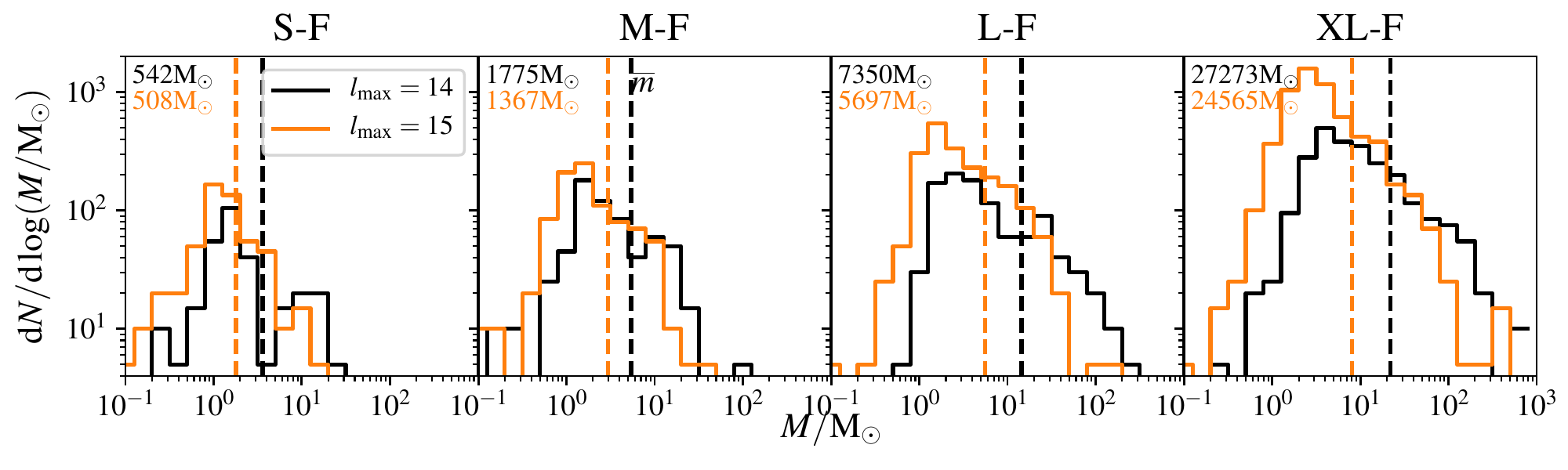}
    \caption{Comparing IMFs from simulations with different resolutions. Fewer massive stars and more lower-mass stars form from simulations with higher resolution.}
    \label{fig:imfres}
\end{figure*}

\revi{
We conduct four extra runs at lower resolution to evaluate the numerical convergence of our simulations. All other simulations have the highest resolution we could afford computationally, and increasing the resolution is unfeasible for the relatively large cloud masses considered in this study. More massive sink particles and fewer low-mass sink particles form in simulations with lower resolution (see Figure~\ref{fig:imfres}). The mean mass of the IMF, represented by the vertical dashed line in the figure, increases by a factor $\sim 2 - 3$ when the spatial resolution halves.  As the resolution increases, while there is no significant change in the total mass in sinks, the mean mass of sinks decreases, suggesting that some of the sinks fragment into smaller sub-clumps.
A model in which the cores form stars with $\sim 30\%$ efficiency and, as we increase the resolution, additional small mass cores form from unused gas at the low-mass end of the CMF, is instead less consistent with our results for two reasons. i) Allowing more of the diffuse gas to form low mass cores, would produce, in some simulations, a total core formation efficiency above unity (in simulations that have f_* > 0.3-0.4). ii) Figure~\ref{fig:imfres} shows that the core formation efficiency, which is $\sim f_*$, is close to being converged. Thus, by increasing the resolution we do not add new cores from the gas, otherwise we would observe an increase of $f_*$, which instead slightly decreases with increasing resolution. Increasing the resolution simply changes the CMF, but the total mass in cores remains nearly the same. For these reasons, we find that the cores-fragmentation model is the most likely, although we cannot rule out alternative scenarios.
}

\subsubsection{Monte-Carlo Numerical Experiments for Fragmentation}\label{sec:MC}

\begin{figure}
    \centering
    \includegraphics[width=\columnwidth]{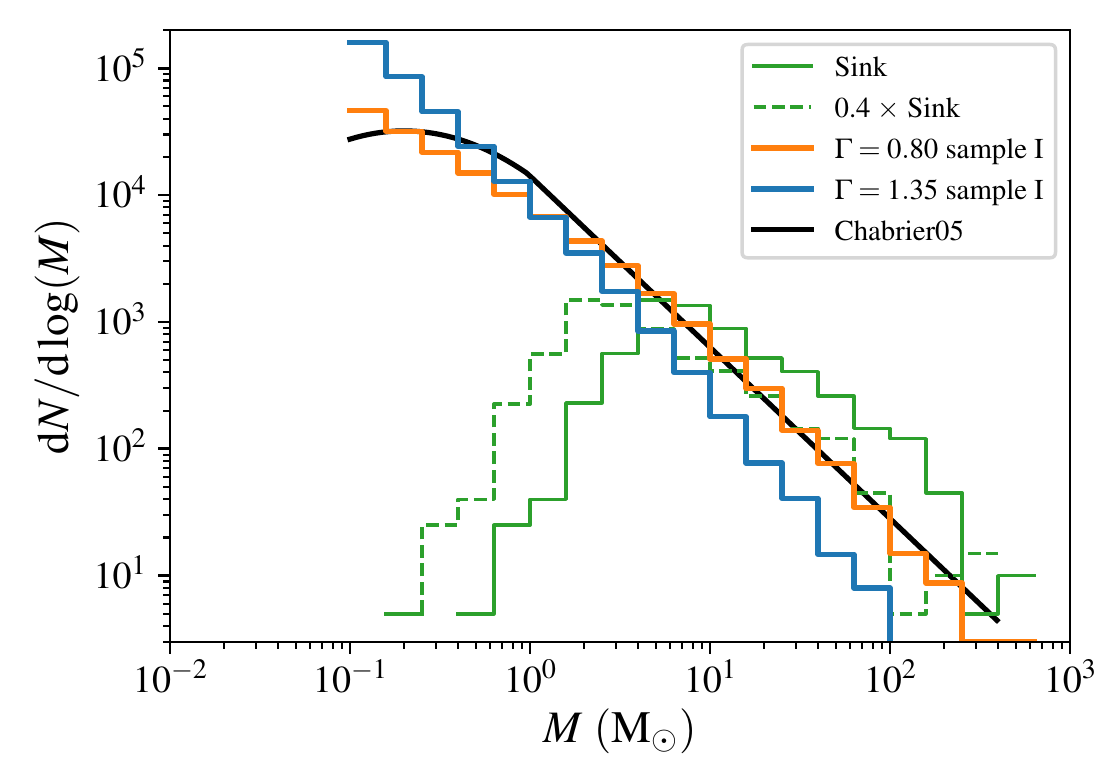}
    \includegraphics[width=\columnwidth]{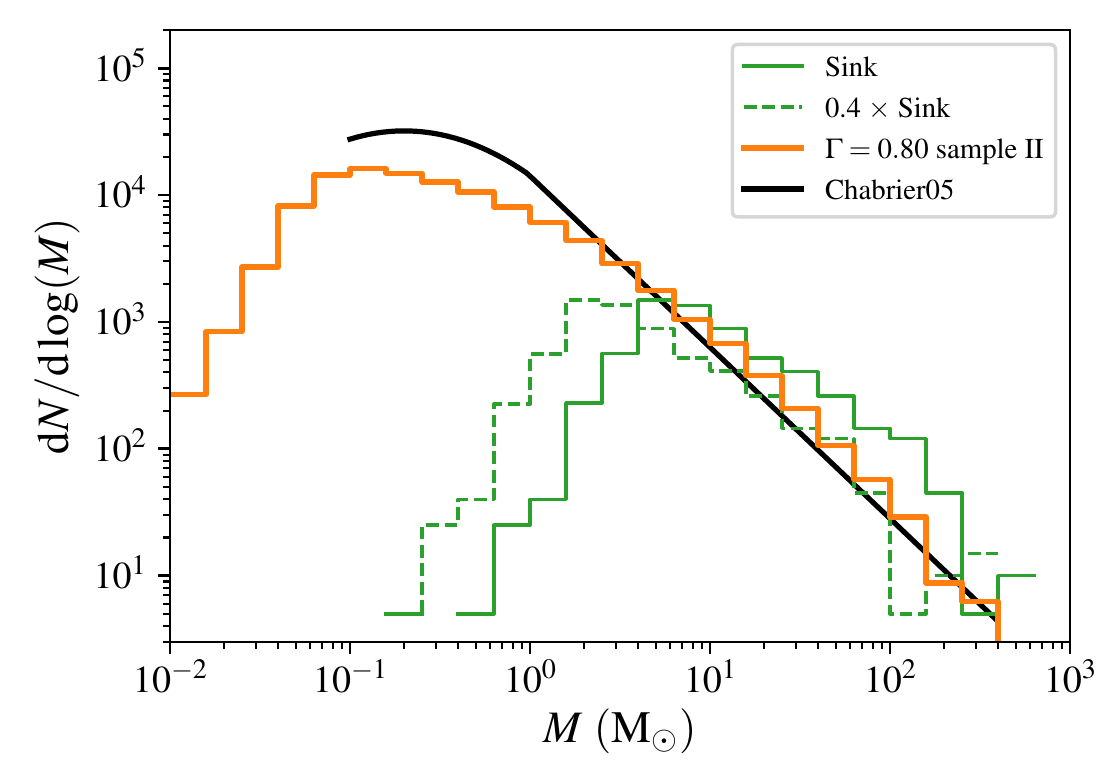}
    \caption{\label{fig:frags}
    Top: Numerical experiments showing the results of fragmenting each sink particle into stars using a power-law PDF with slope $\Gamma$ \revi{for the XL-F simulation}\ignore{, for two simulations: XL-F (top panel) and L-C (bottom panel)}. Similar results are obtained for all the other simulations.
    The \revi{green}\ignore{black} solid and \revi{dashed}\ignore{dash-dotted} histograms are the sink mass function (SMF) and the "shifted" sink mass function (\ie, sinks masses are multiplied by 0.4), respectively. 
    The blue and orange solid histograms show the mass functions of the stars obtained by fragmenting each sink particle into smaller particles using a power-law probability distribution with a slope $\Gamma$ \revi{in the range of $0.1~\msun$ to $m_{sink}$}, as shown in the legend \revi{as `sample I'}.
    Mass-normalised analytic \ignore{Kroupa and }Chabrier IMF are plotted for comparison. A power-law sampling of the sinks with $\Gamma = 0.8$ produces an IMF in very good agreement with a Chabrier IMF over the whole range of star masses.
    \revi{Bottom: Similar to the Top but the lower limit of the sampled masses is set to max$(0.01~\msun, 0.01 m_{sink})$ (sample II).}
    }
\end{figure}

To demonstrate more convincingly that our interpretation is robust, we perform a simple numerical experiment. We assume that each pre-stellar core (sink particle) fragments into smaller sub-units with a power-law mass function (MF) with index $\Gamma$, and with limits on the fragment masses between 0.1~M$_\odot$ and the sink mass. We draw randomly from this distribution until the total mass of the fragments equals the sink mass. We repeat this procedure for all the sinks. 
Such sampling is done 20 times and the average of the bins is taken. 
Figure~\ref{fig:frags} shows the resulting mass function for \revi{the XL-F cloud}\ignore{ two clouds (XLF and LC)}, but we obtain similar results for all the simulations.
The mass function we obtain by fragmenting the sinks is shown as the solid histogram (blue for $\Gamma=1.35$ and orange for $\Gamma=0.8$). The original SMF is shown by the \revi{green}\ignore{black} solid histogram and the shifted mass function is shown by the \revi{dashed}\ignore{dotted} histogram. 
The \revi{black solid curves}\ignore{dashed and dash-dotted lines} show the \ignore{Kroupa and }Chabrier IMF for a cluster with total mass equal to the total mass in sink particles, which are in very good agreement with the mass function of the fragmented pre-stellar cores assuming $\Gamma = 0.8$.

\revi{
This sampling method does not produce a modal mass for the IMF. To address this, we tried another sampling method. However, if we assume that the lower mass limit in the sampling is set to max$(0.01~\msun, 0.01 m_{sink})$, instead of $0.1~\msun$, the resulting IMF has a similar shape to the SMF but peaks at a mass 100 times smaller, resulting in a model mass of the IMF $\sim 0.1~\msun$ (see bottom panel in Figure~\ref{fig:frags}).
}
 
In summary, the fragmentation of the pre-stellar cores into numerous small mass stars, a process which is not captured in our simulations due to limited resolution, explains the deficit of stars with mass below $\sim 1~\msun$ in our simulations with respect to the number expected assuming a \ignore{Kroupa or }Chabrier IMF.

\subsubsection{High-mass Slope of the IMF}
\begin{figure*}
  \includegraphics[width=\textwidth]{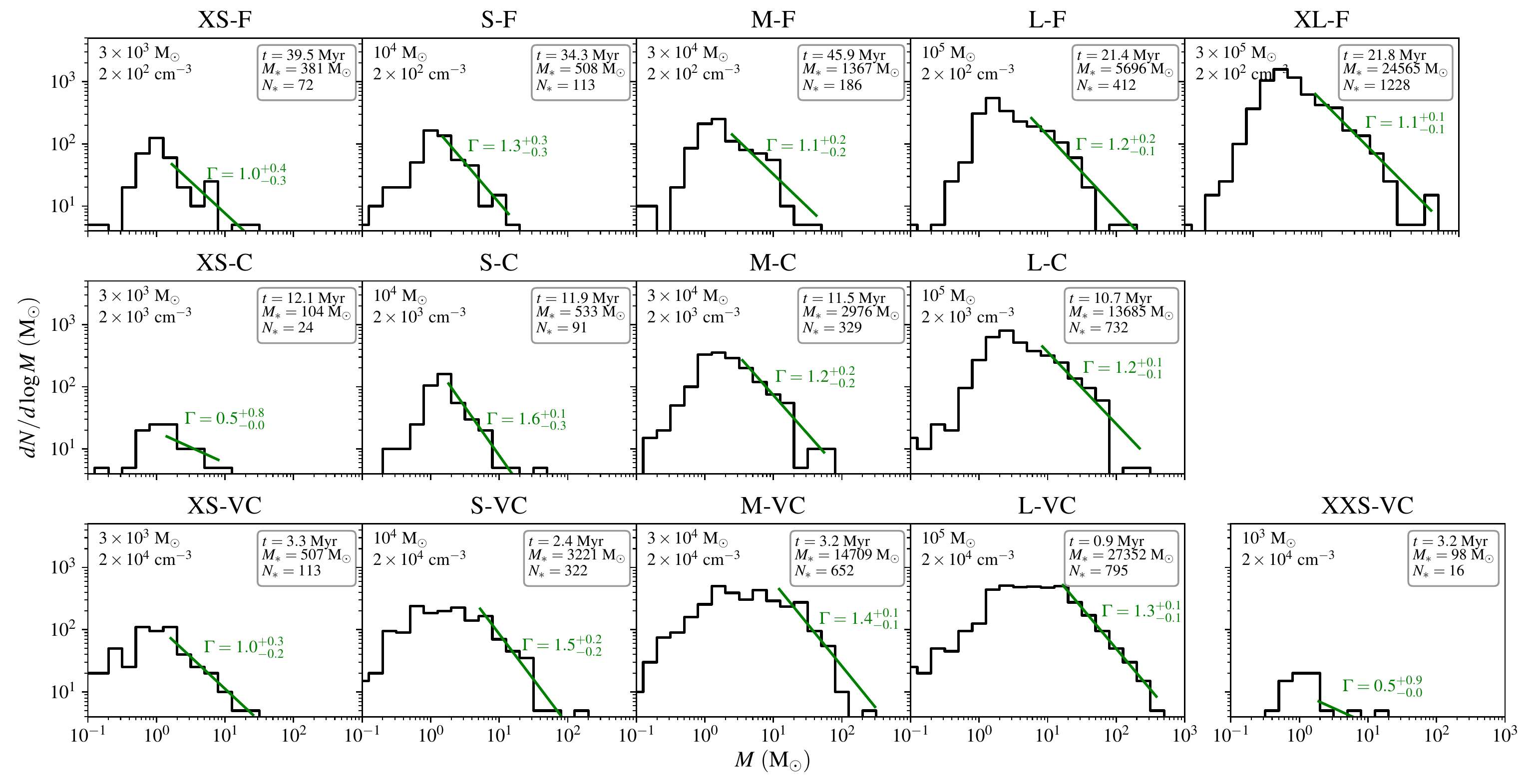}
  \caption{\label{fig:imf} Same as Fig.~\ref{fig:imf_evo}, but showing the IMF at the end of the simulations along with the best fit power-law (solid green lines).
  Only sinks above a critical mass are used for the Bayesian fit. Stars more massive than the critical mass account for $70\%$ of the total cluster mass. The mass range of the sinks used in the fit is also shown as the range of the solid green lines.
  The power-law slopes lie in a range from 1.0 to 1.6 (excluding the simulations that produces less than 50 sink particles), in agreement with the slope of the Salpeter IMF ($\Gamma = 1.35$).
}
\label{fig:imf_fit}
\end{figure*}
 
In this section we quantify more rigorously the slope of the IMF.
In Figure~\ref{fig:imf_fit} the IMF is shown at the end of the simulations when star formation has stopped. The green lines show the best fit power-law at the high-mass end of the IMF using Bayesian inferences as explained below. We do not notice any significant relationship between the high-mass end slope of the IMF and the mass or compactness of the cloud. We notice, however, a flattening of the IMF at 1-10 M$_\odot$ in very compact clouds of high-mass (M$>10^3$~M$_\odot$), which is instead not observed in the fiducial and massive clouds. Since these clouds have the highest star formation efficiency and the strongest radiative feedback, a speculative interpretation would be that we are observing the effect of photo-evaporation of small fragments. When a proto-star is exposed to the ionising flux of a new-born OB star, the disk mass decreases rapidly with time. This may regulate the mass accretion rate through the disk and therefore to the star.

Here are some details of the Bayesian inference of the IMF slope. We assume a power-law slope mass distribution with general form $\md N/\md \log m = A m^{-\Gamma}$ where $A$ is a constant of normalisation and $m_{\rm min} < m < m_{\rm max}$. When the total number of stars is $N_0$, this constant becomes $A = \ln{10}\,N_0 ~ \Gamma / (m_{\rm min}^{-\Gamma} - m_{\rm max}^{-\Gamma})$. The likelihood is proportional to the distribution function,  $\mathcal{N}(\mu_i|\Gamma) = Am_i^{-\Gamma}$, where $\mu_i \equiv \log(m_i)$. To find the most likely $\Gamma$ we calculate the value of $\Gamma$ that maximises the log of the likelihood: 
\begin{align}
    \ln \mathcal{L} &\propto \sum_i \ln \, \mathcal{N}(\mu_i | \Gamma) \nonumber \\
    &= N_0 \left[ \ln \Gamma - \ln(m_{\rm min}^{-\Gamma} - m_{\rm max}^{-\Gamma}) \right] - \ln 10 \, \Gamma \sum_i \mu_i.
\end{align}
Model-independent constants are removed from this equation.
We do the un-binned fitting only to stars with masses above a critical value. This value is chosen somewhat arbitrarily as the point at which the IMF starts to deviate from the Kroupa IMF. In each panel the best fit line is shown as a segment between the critical mass and the maximum stellar mass along with the slope $\Gamma$ and 1-$\sigma$ errors. The errors are calculated as the $16\%$ and the $84\%$ points of the cumulative likelihood for $\Gamma$ between $\Gamma=0.5$ to $1.8$.
The fitted value of $\Gamma$ has a dependence on the critical minimum mass for the points included in the fit, but we find that the values of $\Gamma$ agree with a Kroupa IMF within the 1-$\sigma$ errors in most cases. 
Here, we adopt a critical mass such that particles above this mass account for $70\%$ of the total mass in stars.

\subsubsection{Maximum Stellar Mass in the Cluster}
\label{sec:mass}
\begin{figure}
    \centering
    \includegraphics[width=\columnwidth]{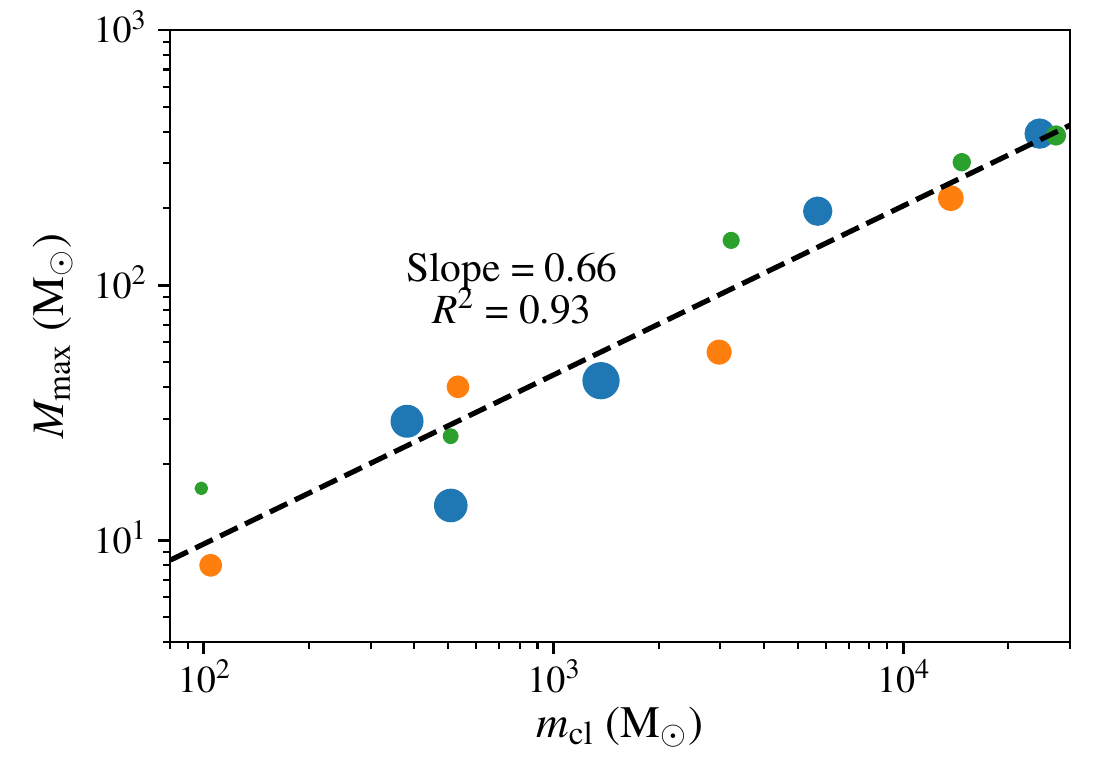}
    \caption{Maximum stellar mass in a cluster v.s. the mass of the star cluster.
    Least-square method is used to fit the data to a power-law with slope $0.66\pm 0.06$.
    The radius of each circle is proportional to the square root of the half-mass effective radius of the cluster and the colour represents the compactness of the cloud: orange for fiducial, blue for compact, and green for very compact.
    }
    \label{fig:max}
\end{figure}

Figure~\ref{fig:max} shows the maximum stellar mass ($M_{\rm max}$) as a function of the mass of the star cluster. The relationship between $M_{\rm max}$, obtained by multiplying the maximum sink mass by 0.4, and the stellar cluster mass, $m_{cl}$, is tight. The best fit power-law is 
\begin{equation}\label{eq:mmax}
    M_{\rm max}/\msun \approx 205 \, m_4^{0.66},
\end{equation}
where $m_4 = m_{cl} / 10^4~\msun$, valid when $m_{cl} \gtrsim 100~\msun$. The relationship is well correlated, with a coefficient of determination R$^2 = 0.93$. 
A power-law relationship between the maximum stellar mass and the cluster mass is consistent with observations, although the observed power-law slope is 0.45 \citep{Larson:1969}, which is slightly flatter than the value found in our simulations. However, the slope we find is in good agreement with numerical studies of star formation in clusters using SPH codes \citep{Bonnell:2003,Bonnell:2004}. We also neglect smaller-scale feedback from protostellar outflows that can reduce the final mass of stars. In addition, it should be kept in mind that the maximum stellar mass here is defined as 0.4 the maximum sink mass, therefore it is possible that the fragmentation of the largest sinks may produce stellar masses systematically smaller than $M_{\rm max}$.

\revi{
Given the uncertainty due to Poisson statistical fluctuations, the SMF appears to be consistent with power-law all the way to the mass bin that is expected to have $\sim 1$ particle in it (the horizontal dashed-dotted lines in Figure~\ref{fig:imf_fit}). Hence we do not have strong evidence for a high-mass truncation of the CMF. We conclude that the CMF, as represented by the SMF, does not have a fundamental upper mass limit below $\sim 1000~\msun$ (the maximum sink mass in all simulations).  Since our simulations have the same initial turbulence field and we have only one random realisation for each set of parameters (mass, and density of the cloud), we are not able to address the question of whether the maximum stellar mass in a cluster is determined by physical \citep{Kroupa:2003} or statistical effects \citep[\eg][]{Fumagalli:2011}. In addition, we use an empirical relationship between sinks mass and massive stars, rather than resolving the fragmentation of sinks into massive stars using a physical model. This also prevents us from drawing robust conclusions about this open question.
}

\subsection{Star Formation Efficiency}
\label{sec:sfe}

\begin{figure}
  \includegraphics[width=\columnwidth]{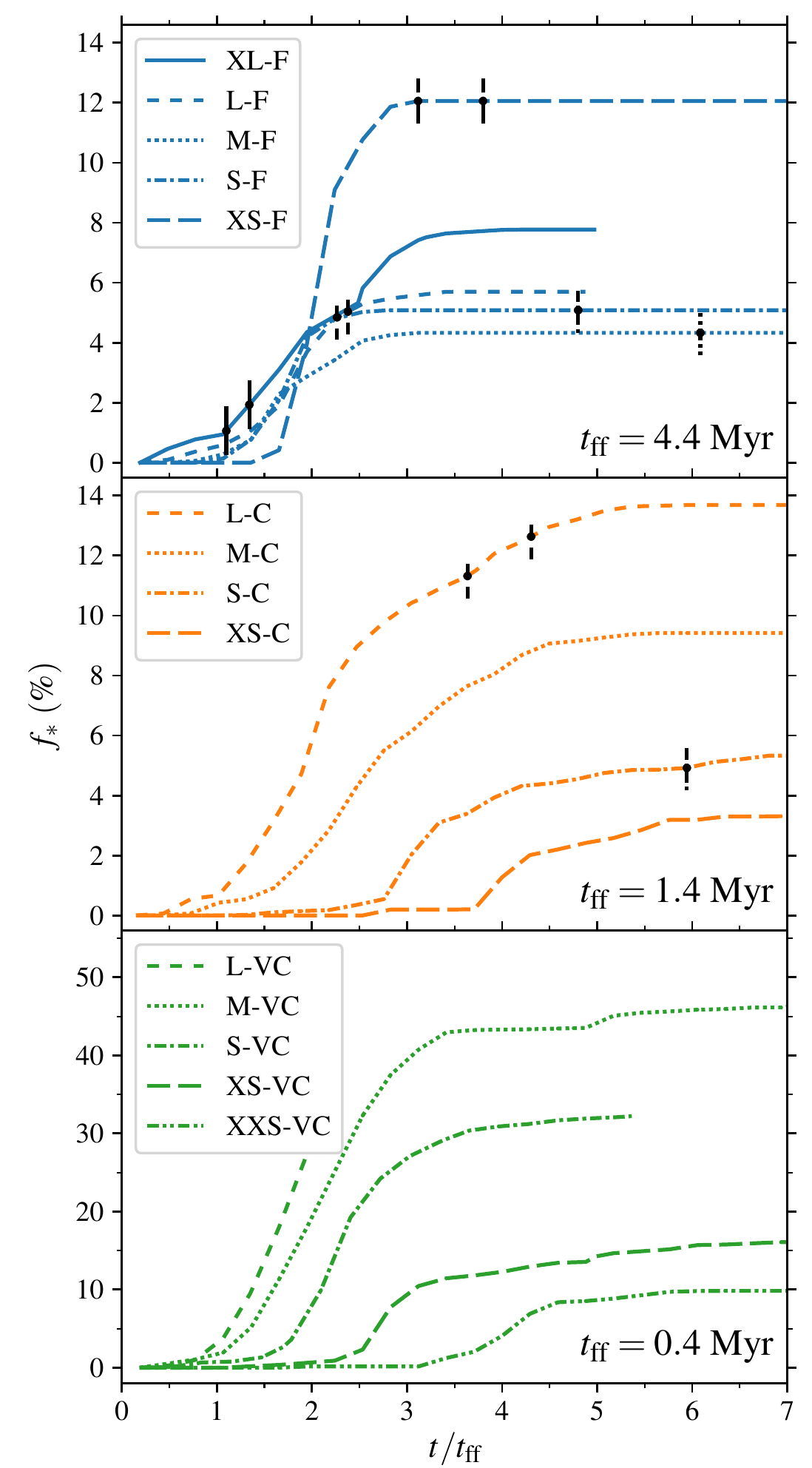}
  \caption{
  Dimensionless star formation efficiency $f_*$\ignore{(the fraction of the mass in gas that is converted into sink particles.)} as a function of the dimensionless time $t/t_{\rm ff}$ for all the simulations shown in Table~\ref{tab:1}. The top, middle, and bottom panels show the fiducial, compact, and very compact clouds, respectively. The black vertical lines indicate the time of the first two SN explosions, if they exist, for each simulation, where the lifetimes of stars are given by \protect\cite{Schaller:1992} fit.
  The duration of the star formation episode is roughly proportional to the sound-crossing time of the cloud (see Sec.~\ref{sec:sflaw}).
  }
  \label{fig:sfe_t}
\end{figure}
We define star formation efficiency (SFE, or $f_{*}$) in our simulated clouds as the fraction of the initial gas mass that is converted into sink particles. Figure~\ref{fig:sfe_t} shows the SFE as a function of time in units of the free-fall time $t_{\rm ff}$ (shown at the top-right of each panel), for the simulations in Table~\ref{tab:2}. The top panel refers to the fiducial clouds, the middle panel to the compact clouds and the bottom panel to the very compact clouds. Lines in each panel refer to different cloud masses as explained by the simulation IDs in the legend. The vertical lines mark the time of the explosion of the first two SNe in the simulation, where the lifetimes of stars are given by \cite{Schaller:1992} fitting functions. As discussed before we do not include mechanical feedback from SNe, but star formation has already stopped or it is mostly terminated before the explosion of the first SN in all simulation but XL-F, \ie the fiducial run with mass $m_{gas}=3.2 \times 10^{5}$~M$_\odot$.

When time is measured in units of the free-fall time, the shape of the SFE curves are qualitatively similar: the SFE increases rapidly with time and peaks at $t\approx 2-3 t_{\rm ff}$.
Generally the total SFE at the end of the simulations increases with increasing cloud mass and with increasing cloud compactness.
\begin{figure}
    \centering
    \includegraphics[width=\columnwidth]{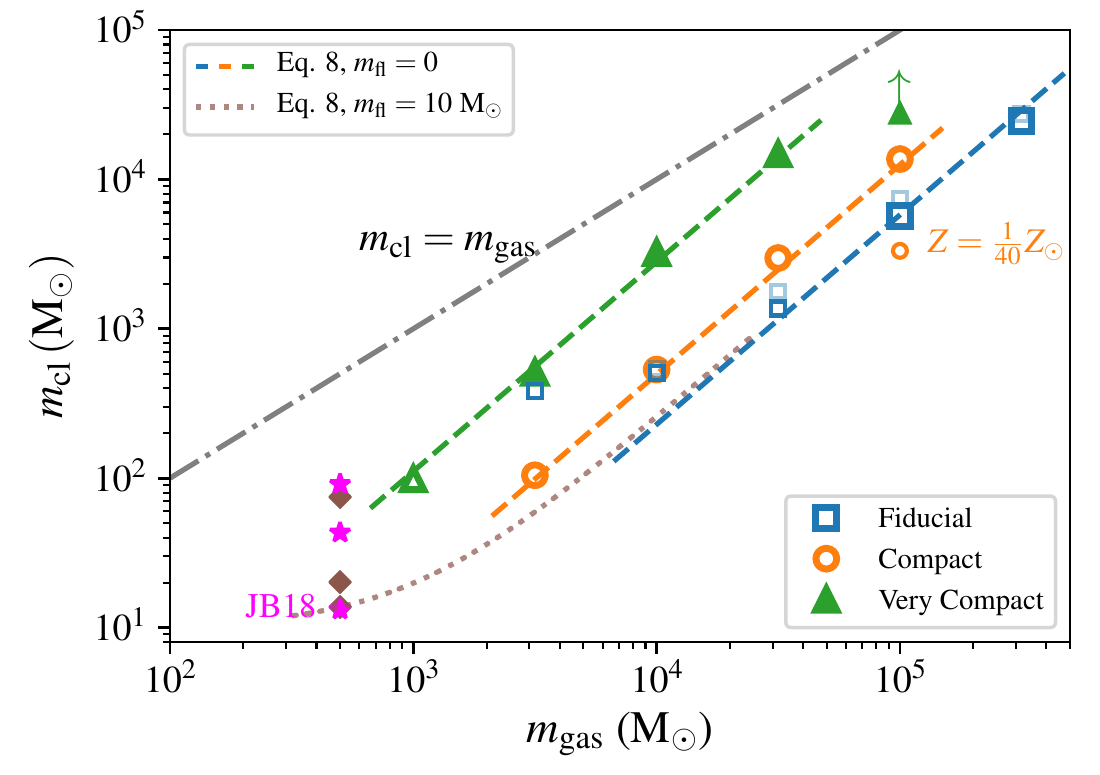}\\
    \includegraphics[width=\columnwidth]{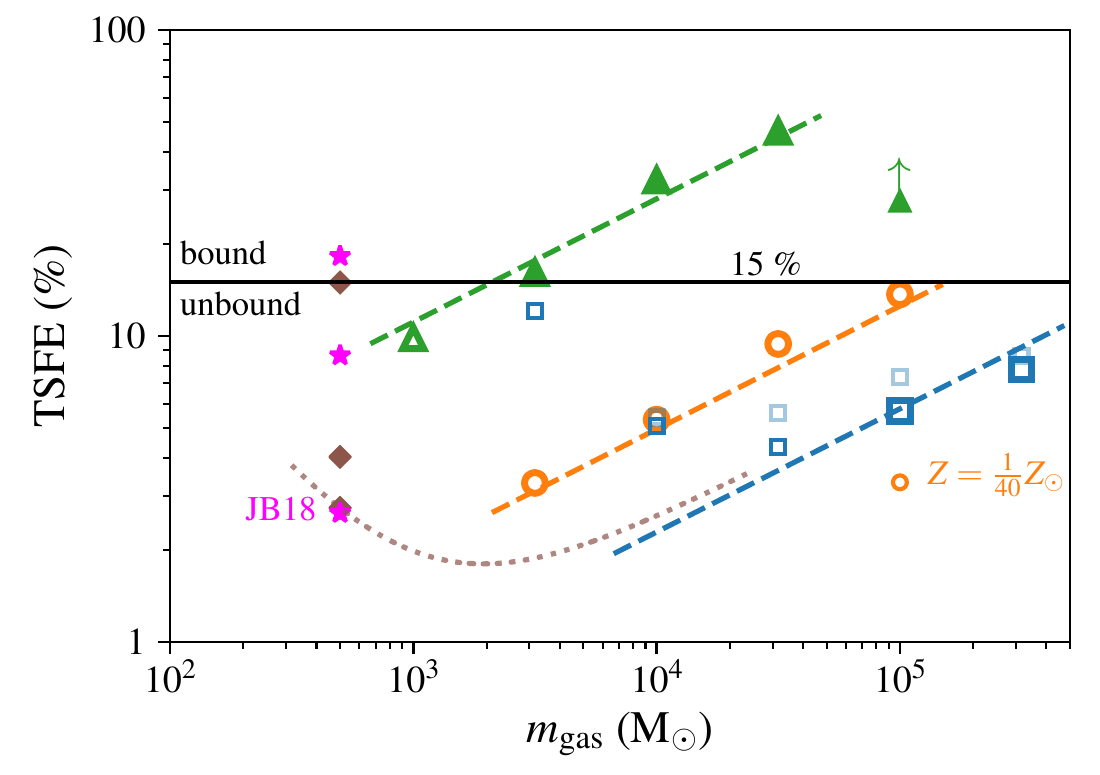}
    \caption{
    ({\it Top.}) Stellar mass of the cluster $m_{cl}$ as a function of the initial mass of the gas cloud ($m_{gas}$) for the set of simulations with different initial cloud densities (see legend). The gray dot-dashed line is plotted as a reference for $100\%$ star formation efficiency. Excluding the 3 fiducial cloud simulations with the lower masses,  
    we observe a clear power-law relation between $m_{\rm cl}$ and $m_{\rm gas}$. We speculate that the minimum cluster mass floor observed for the fiducial clouds data points is due to inefficiency UV stellar feedback due to lack of realistic implementation of low-mass stars feedback in our simulations.
    Indeed the simulations by \protect\cite{Jones:2018}, shown as magenta stars, are in excellent agreement with the extrapolation of out power-law fits as shown by the brown diamonds, assuming Eq.~(\ref{eq:mcl}) and Eq.~(\ref{eq:fstar}) fits with $m_f = 10 \msun$ (see the brown dashed line for our fit to the smallest density of the three Jones18 data points). 
    ({\it Bottom.}) Same as the top panel but showing the total star formation efficiency (TSFE), i.e. the SFE once star formation ends and the cloud is dispersed. The solid horizontal line at $f_*=15\%$ roughly separates clouds that form globular cluster progenitors from open star clusters. 
    }
    \label{fig:m_star}
\end{figure}
This is shown more clearly in Figure~\ref{fig:m_star}.
The top panel in Figure~\ref{fig:m_star} shows the stellar mass of the cluster $m_{\rm cl}$ as a function of the cloud gas mass for the 3 set of simulations with different compactness (as shown in the legend). The smaller open circle with the label $Z=1/40$~Z$_\odot$, shows a compact cloud simulation but with lower gas metallicity (see Section~\ref{sec:met}).
The dot-dashed line shows SFE$=100\%$, while the dashed lines are fits to the simulation results with the following function:
\begin{equation}
  m_{\rm cl} = 200~{\rm M}_{\odot} \cdot \left(\frac{m_{gas}}{10^4~\msun}\right)^{1.4} \left( 1+\frac{\overline{n}_{gas}}{n_{\rm cri}}\right)^{0.91} + m_{fl} \, ,
  \label{eq:mcl}
\end{equation}
where $n_{\rm cri} \approx 10^3 ~\pcc$ is the critical density
and $m_{fl}$ is the mass floor. The dashed lines show the fit assuming $m_{fl}=0$, while the dotted line has $m_{fl}=10$~M$_\odot$.
Equation~(\ref{eq:mcl}) is a good fit to the points when excluding the 3 lowest mass simulations for the fiducial run (shown as smaller sized open squares). The motivation for excluding these 3 simulations from the fits is explained below.

The open symbols show star cluster that become dynamically unbound (\ie, open star clusters), while the solid symbols show star cluster that at the end of the simulations, after most of the gas has been used up for star formation or expelled, remain gravitationally bound (\ie, globular cluster progenitors).

The star symbols show the results of simulations by \cite{Jones:2018} for clouds with mass $m_{gas}=500$ $\msun$ and for
mean densities $\overline{n}_{gas}=\num{3e2}$, $\num{3e3}$, and $\num{3e4}$ cm$^{-3}$,
from bottom to top, respectively. These densities are slightly different from the mean densities in our fiducial, compact and very compact simulations, thus we show as diamonds the corresponding points obtained using our fitting formula in Equation~(\ref{eq:mcl}) with $m_{fl}=10$~M$_\odot$. These simulations do not include feedback by massive stars being very small mass clouds in which the most massive star that forms has is $<10$~M$_\odot$. However, the resolution of these simulations is higher than our simulations and, contrary to our simulations, feedback by IR radiation is included. In addition, these simulation are run using an SPH code. It is interesting to note that despite the different codes and physics included, the results are consistent with the extrapolation of our fitting formulae to low mass clouds if we assume a minimum mass floor for the star cluster mass of $\sim 10$~M$_\odot$.

The bottom panel in Figure~\ref{fig:m_star} is the same as the \revi{top} panel but shows the total star formation efficiency $f_{*, tot}\equiv m_{\rm cl}/m_{\rm gas}$ and the best fit:
\begin{equation}
  f_{*, tot} = 2.0\% \left(\frac{m_{gas}}{10^4~M_\odot}\right)^{0.4} \left( 1+\frac{\overline{n}_{gas}}{n_{\rm cri}}\right)^{0.91}.
  \label{eq:fstar}
\end{equation}

The solid horizontal line at TSFE $\sim 15\%$ roughly separates star clusters that become globular cluster progenitors ($f_*>15\%$) from open star clusters ($f_*<15\%)$. This separation is based on the dynamical state of the cluster at the end of the simulations, but a more detailed analysis of the dynamics of the stellar cluster will be the subject of a followup study.

Let's now address the reason why we excluded the 3 lower mass fiducial simulations from our analysis.
We observe that the star cluster mass in these simulations does not obey a simple power-law relationship with the initial gas mass of the molecular cloud. The discrepancy does not appear to be a convergence issue due to insufficient resolution, as confirmed by the lower-resolution simulations (shown as lighter colour small squares), but rather lack of the necessary physics for self-regulation feedback.
This can be understood inspecting Figure~\ref{fig:max} which relates the mass of massive stars to the cloud gas mass. The low TSFE of the diffuse clouds in combination with the small cloud gas mass produces stellar masses below $10^2$~M$_\odot$, which corresponds to a maximum stellar mass $M_{max}<10$~M$_\odot$. Such stars do not produce significant quantities of ionising UV radiation, therefore the cloud can continue to form stars. This is due to our neglecting feedback mechanisms from lower mass stars. This requirement for stars that produce ionising radiation to disperse the cloud leads to a minimum cluster mass floor $m_{\rm cl} \sim 300$~M$_\odot$, much larger than the $\sim 10$~M$_\odot$ floor which is a good fit to the simulations of \cite{Jones:2018}. 

This large mass floor is not evident in the compact and very compact clouds: if it exists, it must be at masses $m_{\rm cl} <100$~M$_\odot$.
The reason for this apparent inconsistency is not fully understood, but it appears to be related to the smaller ratio of the crossing to free-fall time for the fiducial cloud when compared to the more compact clouds (Sec.~\ref{sec:sflaw}).
We offer the following hypothesis: Inspecting the middle and bottom panels in Fig.~\ref{fig:sfe_t}, we observe a longer delay for onset of star formation in the small mass clouds for the compact and very compact runs, which is not observed in the fiducial runs. This can be understood in terms of the necessary number of crossing times required by the supersonic turbulence to create dense clumps for star formation (with $n>n_{\rm sink}$). In the fiducial cloud this enhancement of the density due to supersonic turbulence is faster when compared to the free-fall timescale, hence the steeper rise of $f_*$ as a function of time. When feedback from massive stars is absent due to random sampling of a small mass stellar cluster, this rapid increase of $f_*$ can lead to significant overshooting of star formation above the threshold expected from self-regulation. This overshooting does not happen, or is milder, for more compact clouds in which $f_*$ increases with $t/t_{ff}$ more slowly. The existence of a minimum cluster mass floor, however, should eventually become evident also in more compact clouds when decreasing further the initial cloud masses.

\revi{
We observe a power-law relation between the mass of the cloud and the mass of the star cluster.
\cite{Howard:2018a} find that the stellar mass of the most massive cluster that forms from a molecular cloud has a power-law dependence on the mass of the cloud with an exponent of $0.78$. 
In our work, this relation, taking all sink particles as the cluster, 
has an exponent of 1.4 (Equation~\ref{eq:mcl}). By multiplying it with 
the exponent of the $M_{max}$-$m_{cl}$ relation, 0.66  (Equation~\ref{eq:mmax}), we get an exponent of 0.92. Similar to \cite{Howard:2018a}, our work suggests that young massive star clusters are natural extensions of low-mass cluster formation.
}

\revi{
Our results (Figure~\ref{fig:m_star}, or Table~\ref{tab:2}) are in good agreement with \cite{Kim:2018}, who find that the TSFE depends primarily on the initial gas surface density, such that the TSFE increases from $4\%$ to $51\%$ as $\Sigma$ increases from 13 to 1300 $\msun ~ {\rm pc}^{-2}$.
}

To summarise, we believe that the increase in TSFE observed for the fiducial simulations with masses $m_{\rm gas} \le 10^4$~M$_\odot$ is unphysical, meaning that it is due to missing feedback processes in our simulations. When the most massive star has mass $M<10$~M$_\odot$, IR radiation feedback or proto-stellar jets feedback should be included in the simulation. In all the other simulations UV feedback by massive stars is likely the dominant feedback at play; therefore these simulations incorporate the relevant physics for the formation of realistic star clusters.

\begin{figure*}
    \centering
    \includegraphics[width=6in]{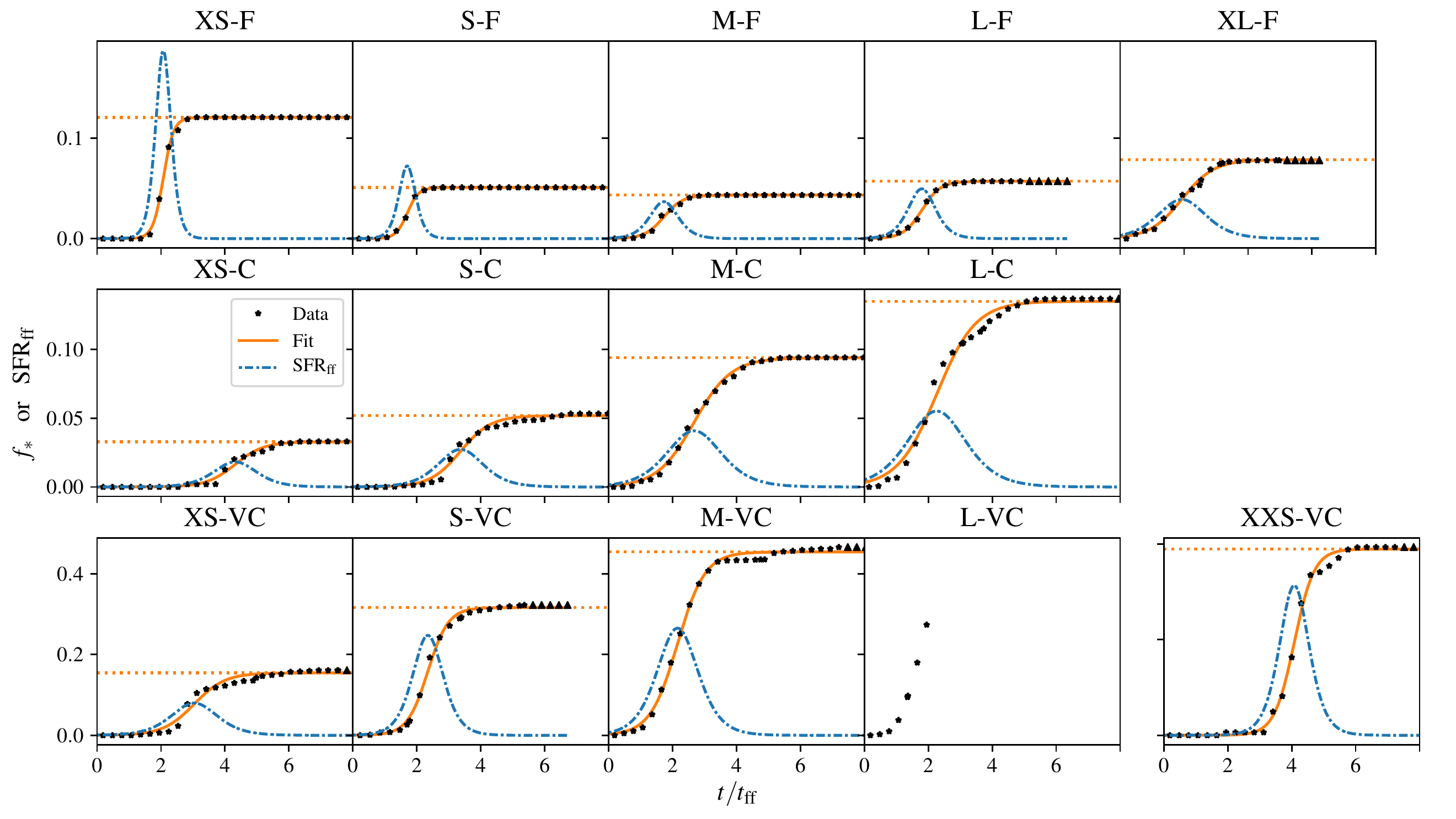}
    \caption{Dimensionless star formation efficiency ($f_* = m_* / m_{gas}$) and dimensionless star formation rate per free-fall time (SFR$_{ff}=df_*/d\tau$) as a function of dimensionless time $\tau=t/t_{\rm ff}$ for the simulations in Table~\ref{tab:1}. The points show $f_*$ as a function of time from the simulations, the solid orange line shows a fit to $f_*(\tau)$ using Fermi function (Eq.~\ref{eq:fermi}), and the solid blue line shows SFR$_{ff}$ using the fit formula.
    The Fermi function is a good fit to the data, and from it we can calculate the peak star formation rate and star formation time (shown in Fig.~\ref{fig:dtau}).
    }\label{fig:sfefit}
\end{figure*}

\begin{figure}
    \centering
    \includegraphics[width=3in]{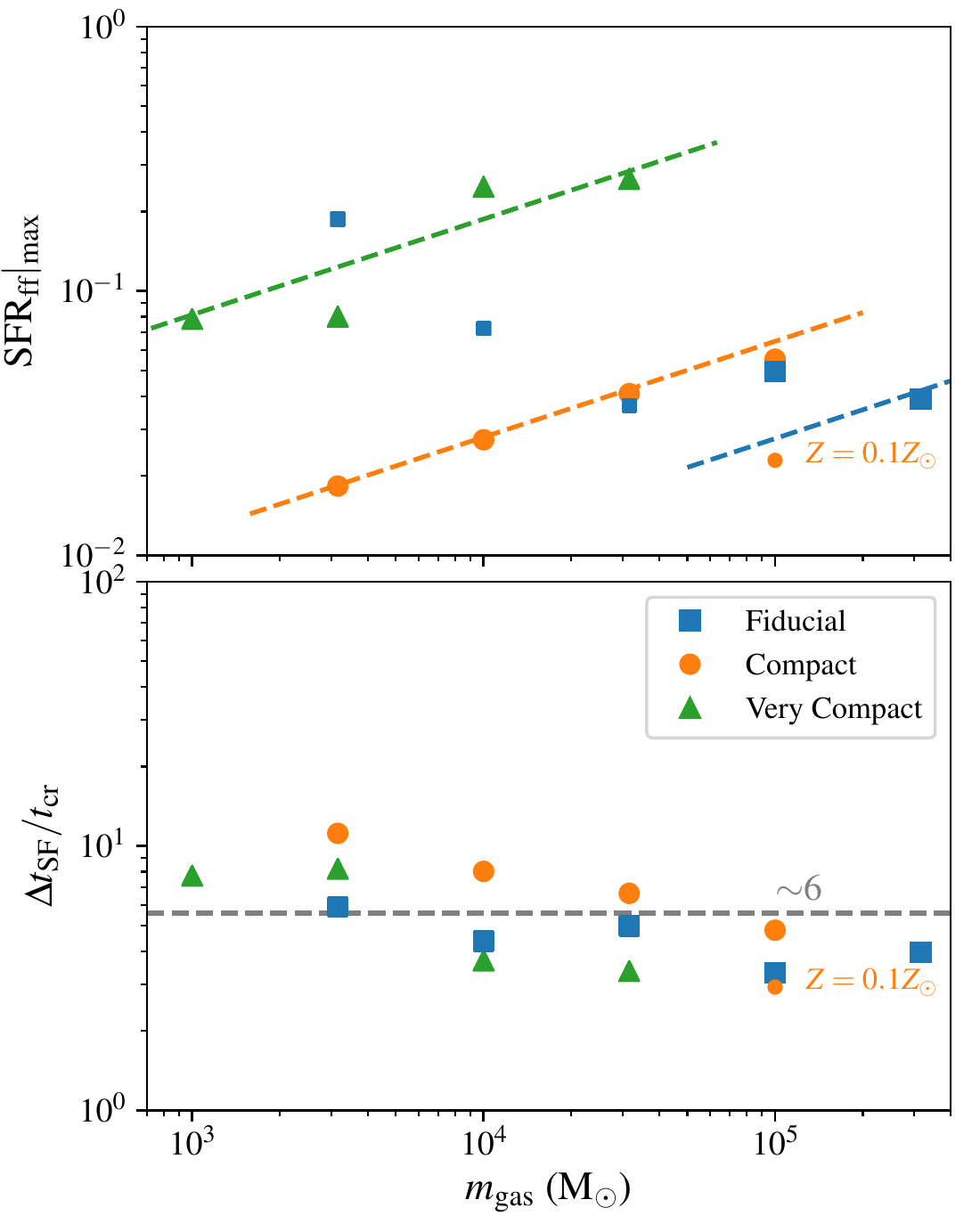}
    \caption{
    {(\it Top)}: Maximum dimensionless star formation rate per free-fall time, SFR$_{ff}|_{max}=\md f_* / \md \tau |_{max}$, where $\tau = t/t_{ff}$ v.s. gas mass of the cloud. 
    The dashed lines show a power-law fit to the data (see Eq.~\ref{eq:sfrff}). The smaller squares are data points not used for the fit because of the lack of a realistic feedback loop in these simulations.
    {(\it Bottom)}: The ratio of star-formation time $\Delta t_{\rm SF}$ to sound-crossing time $t_{\rm cr} = r_{\rm gas} / c_s$, where $c_s = 10 ~\si{km.s^{-1}}$. This ratio is close to a constant (the gray dashed line).
    Over-pressured \HII regions require approximately 6 crossing times to suppress star formation.
    }\label{fig:dtau}
\end{figure}

\subsection{Star formation law in molecular clouds}
\label{sec:sflaw}
Next, we ask the question of what is the physical interpretation of
the empirical relationship we derived for the star formation
efficiency as a function of cloud mass and compactness. To answer this
question we first fit the SFE $f_{*}(\tau)$ with an analytic function, where $\tau \equiv t/\tff$, in
order to minimise the stochastic noise of the simulations. The
$f_{*}(\tau)$ has a shape that can be fit by an $\arctan$ function or the
Fermi function:
\begin{equation}
\label{eq:fermi}
f_F(\tau) = \frac{f_0}{e^{-(\tau-\tau_0)/\Delta\tau} + 1}.
\end{equation}
Both fits give similar results for the
purpose of interpreting $f_*(\tau)$.  In Figure~\ref{fig:sfefit} we
show the fit to $f_{*}(\tau)$ using the Fermi function $f_F$ (orange
solid curves) and its time derivative (blue curves), or 
the dimensionless SFR per free-fall time,
SFR$_{ff} \equiv \md f_*/\md\tau \approx \md f_F / \md \tau$.  The
fits are a good approximations to the data points from the simulations
(solid points), except for a few clouds where $f_{*}(\tau)$ has a pit
near the end of the star formation process.

The value of the peak of SFR$_{ff}$ has a weak dependence on the cloud mass
(see top panel in Figure~\ref{fig:dtau}) and a stronger dependence on
the cloud mean density. We fit the SFR$_{ff}|_{\rm max}$ with a power-law similar to Eq.~(\ref{eq:fstar}):
\begin{equation}
\label{eq:sfrff}
{\rm SFR}_{ff} |_{max} 
\approx 1.1\% \left(\frac{m_{gas}}{10^4M_{\odot}}\right)^{0.36} \left(1+\frac{\overline{n}_{\rm gas}}{n_{cri}}\right)^{\alpha_f}        
\end{equation}
where $\alpha_f \approx 1.0$ and $n_{cri}$ is the same critical density as in Eq.~(\ref{eq:mcl})\footnote{The value of $\alpha_f$ is somewhat correlated with $n_{\rm cri}$. We sample a sequence of $n_{\rm cri}$ for which we obtain a good fit and find that for $n_{\rm cri}$ in the range $ \sim 400$ -- $1600 \pcc$, the corresponding $\alpha_{\rm f}$ is in the range $0.85 \-- 1.1$.}. 
The duration of the star formation burst in units of $\tff$, $\Delta \tau_{SF}$, is proportional to the width of the SFR$_{ff}$ shown as the blue lines in
Fig.~\ref{fig:sfefit}. 
The function $\md f_F / \md \tau$ has a peak value 
$f_0/4\Delta\tau$ and a full-width half-maximum
$3.526 \Delta\tau$. We define $\Delta \tau_{SF} \equiv
4 \Delta{\tau} $ so that 
\begin{equation}
\label{eq:ftot2}
f_{*, \rm tot} \approx f_0 = \frac{\md f_F}{\md \tau}|_{\rm max} \times \Delta \tau_{SF}.
\end{equation}
\ignore{(Fig.~\ref{fig:dtau})} Inspecting Fig.~\ref{fig:sfefit} we see that $\Delta \tau_{SF}$ increases with the cloud mass, and appears
to be proportional to the dimensionless sound crossing time of the cloud. Here we define the sound crossing time, $t_{\rm cr}$, as the ratio of the time it takes for a sound wave with $c_s=10$~km/s to cross the cloud radius. Similarly to the dimensionless $\Delta \tau_{\rm SF}$, we define $\tau_{cr} \equiv t_{\rm cr}/\tff$, where the free-fall time is defined at the cloud's mean density. We find that $\Delta \tau_{\rm SF} / \tau_{\rm cr} = \Delta t_{\rm SF}/t_{\rm cr} \approx 6 $ (the horizontal line in the bottom panel of Fig.~\ref{fig:dtau}).
This results makes physical sense because the feedback mechanism stops star formation by creating
over-pressured \HII regions which require a constant number of crossing
times to expel the gas. 

Since 
$t_{\rm cr} \propto r_{\rm gas} \propto (m_{gas}/\overline{n})^{1/3}$, 
we have
$\Delta \tau_{SF} \propto t_{cr} / \tff \propto m_{gas}^{1/3} \overline{n}^{1/6}$.
From Equation~(\ref{eq:ftot2}) we derive
$f_{*,\rm tot} \propto m_{gas}^{\;\;0.69} \overline{n}_{gas}^{\;\;0.17}(1+\overline{n}_{gas}/n_{\rm cri})^{1.0}$,
which is in good agreement with Eq.~(\ref{eq:fstar}) for
$\overline{n}>n_{\rm cri}$. The agreement can be improved further by considering a more accurate fit to $\tau_{SF}/\tau_{cr}$ rather than assuming a constant value $\sim 6$. Namely, considering the weak dependence of the star formation timescale on the cloud mass and density: $\Delta \tau_{SF}/\tau_{cr} \propto m_{gas}^{\;\;-0.3} \overline{n}_{gas}^{\;\;-0.2}$. 

From the analysis and interpretation of these results we can thus derive a star formation law in molecular clouds that can be used as a more accurate sub-grid recipe in cosmological simulations that resolve the molecular cloud phase. Assuming a constant mean volume for the cloud we have $f_*\equiv m_*/m_{\rm gas}\approx \rho_*/\rho_{\rm gas}$. Therefore, assuming $\rho_{\rm gas}={\rm const}$ (\ie, assuming $f_* \ll 1$) during the episode of star formation, which has a duration $\Delta t_{SF}$, we have SFR$_{ff}|_{max} \equiv df_*/d\tau|_{\rm max} \approx d\rho_*/dt|_{\rm max}(t_{\rm ff}/\rho_{\rm gas})$, which implies:
\begin{eqnarray}
\frac{\dif \rho_*}{\dif t} = \epsilon \left(\frac{m_{gas}}{10^4M_{\odot}}\right)^{0.36} \left(1+\frac{\overline \rho_{\rm gas}}{\rho_{\rm cri}}\right)^{1.0} 
\frac{\overline{\rho}_{\rm gas}}{t_{\rm ff}} \propto (\overline{\rho}_{\rm gas})^{2.5}, \\ \textrm{if } 
\overline{n}_{gas} > n_{cri} \approx 10^3 \pcc 
\nonumber
\end{eqnarray}
with $\epsilon=1.1\%$ for solar metallicity and $\epsilon=0.36\%$ for $Z<0.1$~Z$_\odot$ (see \S~\ref{sec:met}). 
A star formation law $d\rho_*/dt \propto \rho_{\rm gas}^n$ with $n=1$ or $n=1.5$ is most often used as a sub-grid star formation recipe in cosmological simulations. 
Therefore we suggest that a steeper power-law index $n \sim 2.5$ is a better description of the star formation rate at densities typical of molecular clouds in high-redshift galaxies. This theoretical result can, in principle, be tested against observations of young stellar clusters in our galaxy.

\revi{
\cite{Krumholz:2012a} suggests that star formation law is universal in which the star formation rate is $\sim 1.5\%$ of the molecular gas mass per local free-fall time.
Eq.~\ref{eq:sfrff} results in SFR$_{ff} \approx 1\%-2\%$ at $\overline n_{gas} \lesssim 10^3 \si{cm^{-3}}$, in agreement with this work for local molecular clouds. However, \cite{Krumholz:2012a} finds this universal value also for high-redshift galaxies but averaged over the whole galaxy. We find that SFR$_{ff}$ can be as large as $\sim 10\%$ for more compact clouds typical of high-redshift galaxies, and/or more massive clouds (see also the top panel of Fig. 12). 
A direct comparison to Krumholz's results is not trivial for the galaxy as a whole, as it depends on modelling the multi-phase ISM of high-z galaxies.
}

\subsection{Effects of Lowering the Gas Metallicity}
\label{sec:met}
\begin{figure*}
    \centering
    \includegraphics[width=\textwidth]{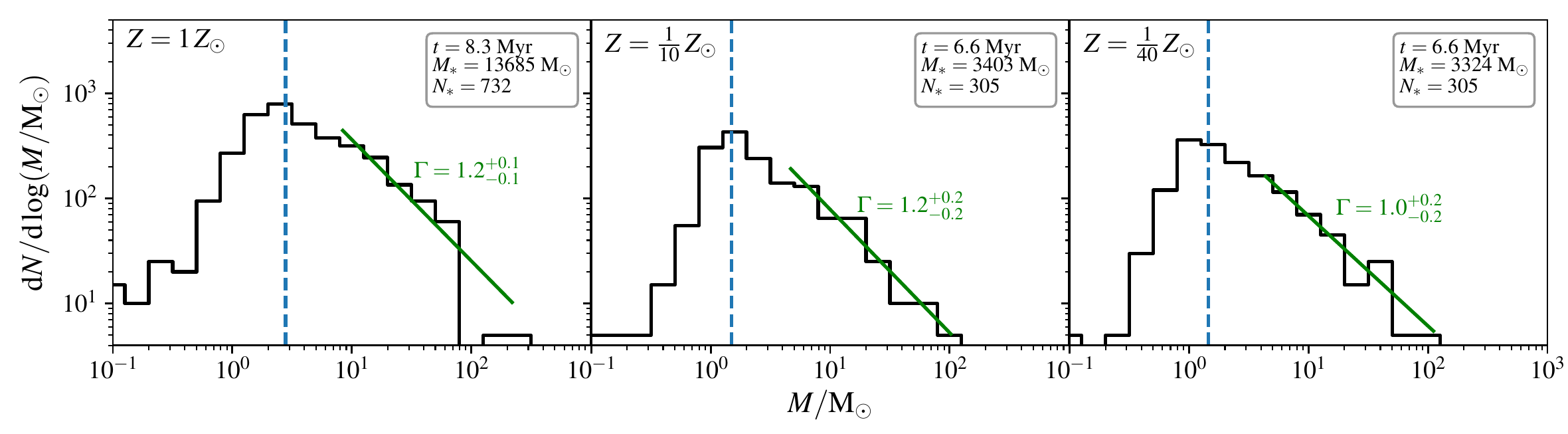}
    \caption{Same as Fig.~\ref{fig:imf} but for the L-C cloud with various metallicities. The metallicities are marked at the top-left corner. We see no significant difference on the shape of the IMF for clouds with different metallicities.}
    \label{fig:imf_met}
\end{figure*}
\begin{figure}
    \centering
    \includegraphics[width=\columnwidth]{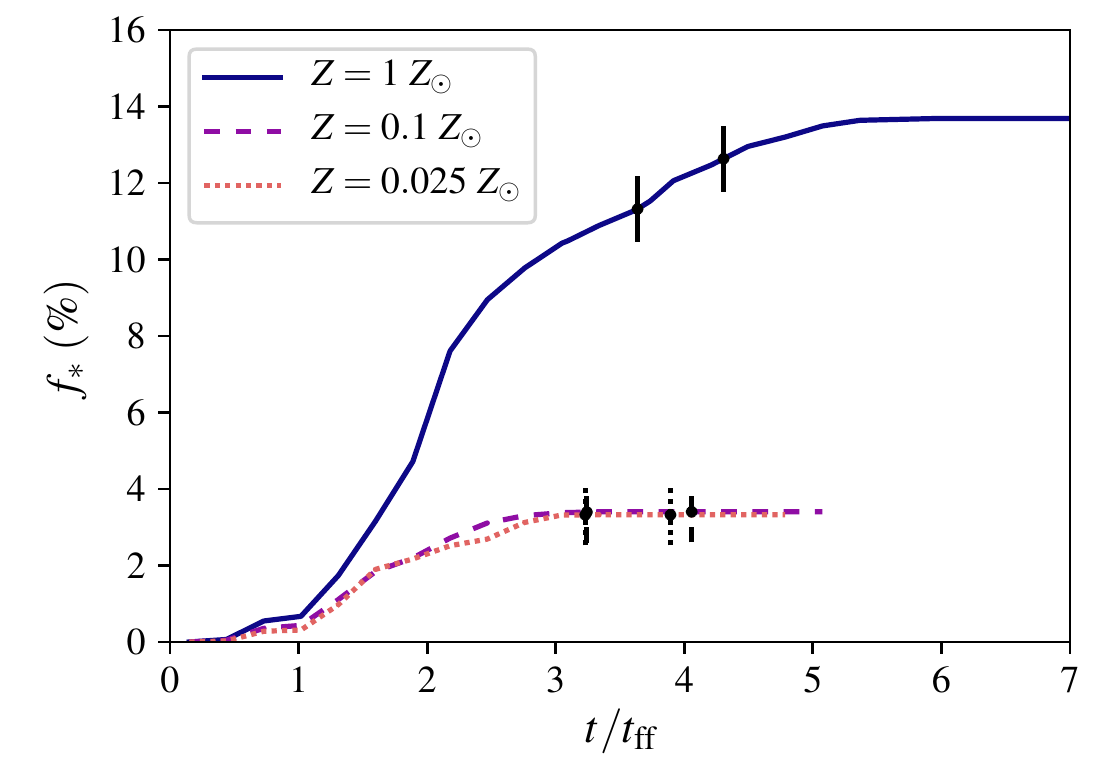}
    \caption{Same as Fig.~\ref{fig:sfe_t} but for the L-C cloud with different gas metallicities, as shown in the legend. }
    \label{fig:sfe_met}
\end{figure}
The set of compact and very compact molecular clouds we have analysed are meant to represent clouds typical of the ISM in dwarf galaxies forming at high-redshift. However, we also know that the gas metallicity in these dwarf galaxies is less than solar. In order to keep the parameter study consistent we have not changed the gas metallicity in the compact and very compact clouds, but in this section we briefly test the influence of gas metallicity [Fe/H] on the star formation rate and IMF. In our simulations, changing the gas metallicty affects the cooling of the gas (see Section~\ref{sec:cooling}).

Figure~\ref{fig:imf_met} is the same as Figure~\ref{fig:imf} but for the LC clouds with metallicity $Z=1 ~Z_\odot$, $0.1 ~Z_\odot$, and $0.025 ~Z_\odot$.
The shape of the IMF is not affected by the gas metallicity. Only the normalisation of the IMF is influenced because of the lower SFE in the low-metallicity simulations. \revi{This is in agreement result of previous theoretical works \cite[\eg][]{Myers:2011,Bate:2014}.}

Lower metallicity translates into lower cooling rates, which should result in lower efficiency of star formation. 
Figure~\ref{fig:sfe_met} shows $f_*$ as a function of time 
for the large compact cloud (LC) with intermediate (0.1 $Z_\odot$) and low (0.025 $Z_\odot$) metallicity. The effect of lowering the metallicity by a factor of ten, from $Z=1$~$Z_\odot$ to $Z=0.1$~$Z_\odot$ is to lower $f_*$ at the end of the simulation by roughly a factor of 5. But lowering further the metallicity from $Z=0.1$~$Z_\odot$ to $Z=0.025$~$Z_\odot$ does not change $f_*$, suggesting that $f_*$ decreases almost linearly with the metallicity from solar to $Z=0.2$~$Z_\odot$, but this effect saturates when further lowering the metallicity.
The SFE decreases mainly because the peak SFR decreases by roughly a factor of 3 with decreasing metallicity, while the duration of the star formation episode is nearly unchanged (see small circles in Fig.~\ref{fig:dtau}).

\begin{figure*}
    \centering
    \includegraphics[width=.32\textwidth]{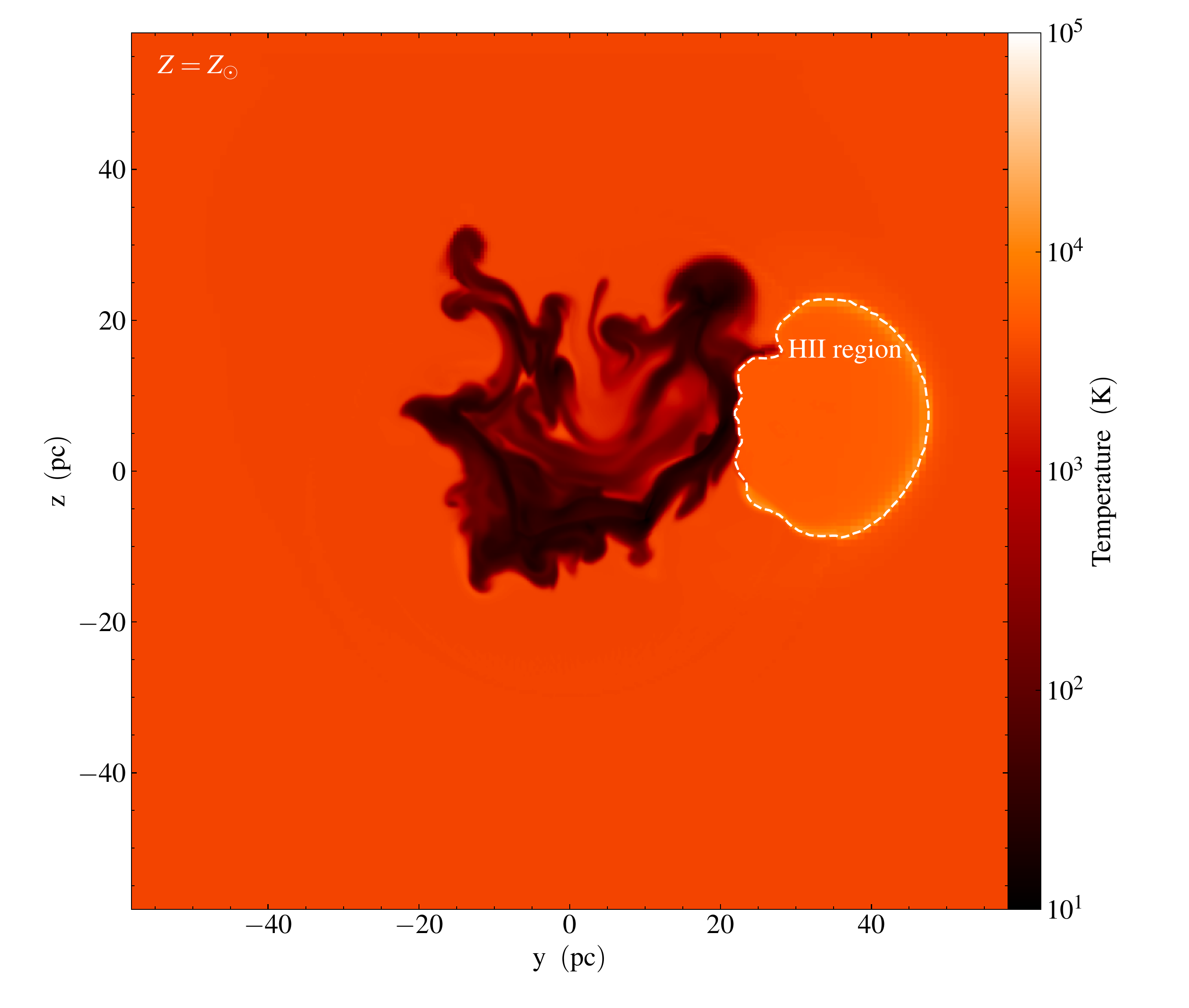}
    \includegraphics[width=.32\textwidth]{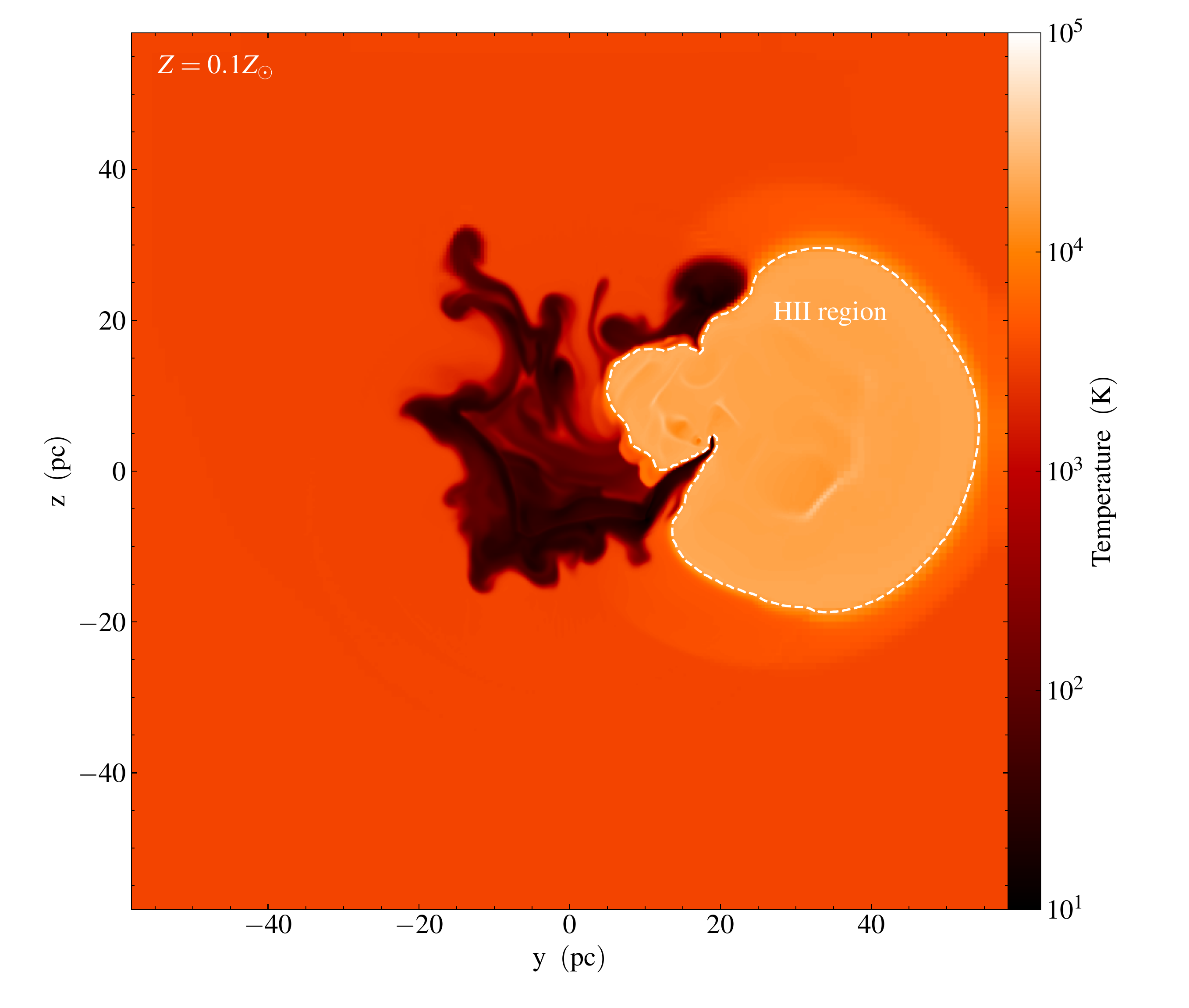}
    \includegraphics[width=.32\textwidth]{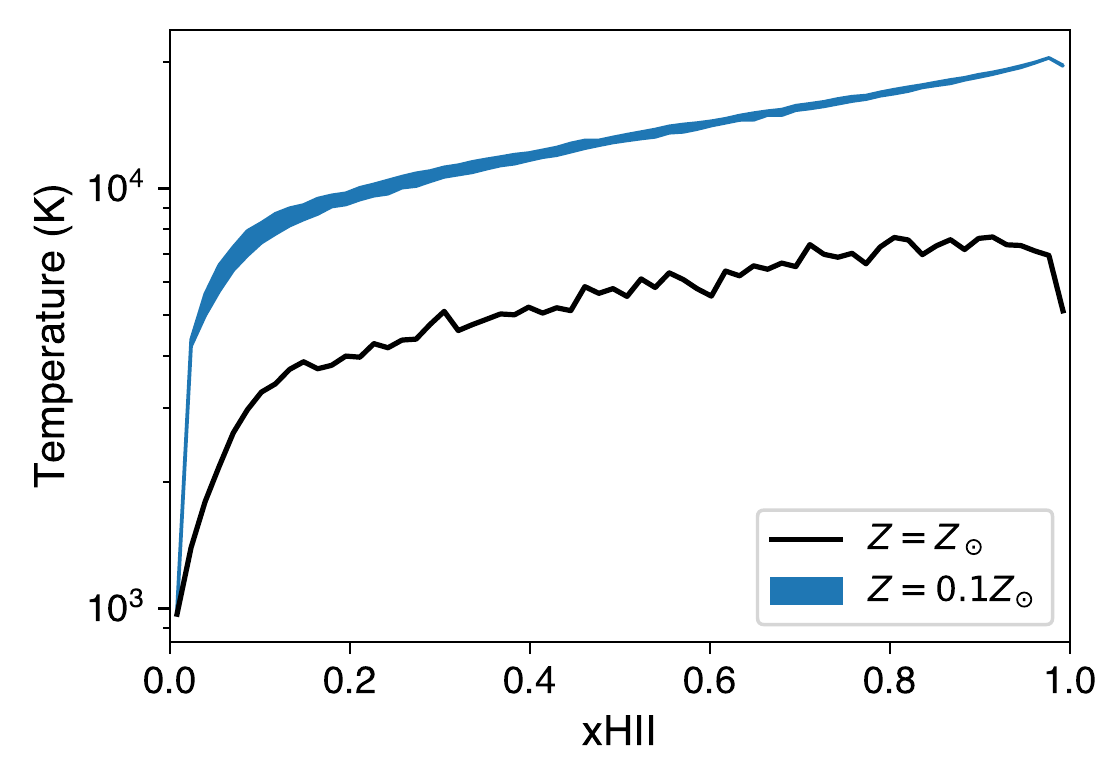}
    \caption{
    {\it (Left and Middle)}. Slice plots of the gas temperature from simulations with metallicities $Z=1Z_\odot$ (left) and $Z=0.1Z_\odot$ (middle). The snapshots from these two simulations are chosen to be nearly at the same evolutionary stage. We observe a factor a $\sim 3$ increase in temperature (and thermal pressure) within the \HII region as the metallicity of the gas is decreases from solar metallicity to a tenth of it.
    Right: Phase plot of gas temperature vs hydrogen ionising fraction for the \HII regions shown in the left and middle panels. The blue shaded area refers to the $Z=0.1Z_\odot$ simulation for a small range of evolutionary times around the time of the $Z=1 Z_\odot$ snapshot (shown as black line).}
    \label{fig:T_vs_xHII}
\end{figure*}
\revi{
In order to better understand what is causing a decrease of the SFR at lower metallicity, we have analysed the density and temperature structure of these two simulations.
We found that lowering the metallicity causes the temperature and the thermal pressure inside \HII regions to increase by roughly a factor of $3$, as shown in Figure~\ref{fig:T_vs_xHII}. This result is in agreement with observations and theoretical models of \HII regions. 
The strength of feedback, due to the increase of thermal pressure inside the \HII regions, is therefore stronger at lower metallicity, resulting in a lower star formation efficiency. This result on the effect of the gas metallicity goes in the opposite direction of what found by \cite{Howard:2018a}. In their work, lowering the metallicity of the gas cloud reduces the opacity of the gas to radiation and results in higher gas accretion which leads to an increase of the total star formation efficiency. However, this can be understood because in their simulations the dominant feedback mechanism is IR radiation pressure while, contrary to our work, UV feedback does not play a major role. However, their simulations describe more massive clouds and have much lower resolution than the simulations in our work.}

\section{Summary and Conclusions}
\label{sec:summary}

In this paper, the first of a series, we present a large set of radiation-magneto-hydrodynamic simulations of star formation in self-gravitating, turbulent molecular clouds.
The initial conditions for the clouds are isothermal spheres initially close to virial equilibrium, being supported by turbulent motions. 

We model the formation of individual massive stars, replacing self-gravitating clumps that are collapsing below the resolution of the simulations with sink particles, which represent individual massive stars, therefore including their UV radiation feedback self-consistently. We consider a grid of simulations varying the cloud masses between $m_{\rm gas}=10^3$~M$_\odot$ to $3 \times 10^5$~M$_\odot$. Depending on the cloud mass, we resolve scales between 200~AU to 2000~AU. In addition, we consider three compactness for the molecular clouds. The fiducial clouds have gas mean number densities typical of those observed in the local universe ($\overline{n}_{\rm gas} = 1.8 \times 10^2$~cm$^{-3}$). Compact ($\overline{n}_{\rm gas} = 1.8 \times 10^3$~cm$^{-3}$) and very compact ($\overline{n}_{\rm gas}=1.8 \times 10^4$~cm$^{-3}$) clouds represent clouds expected to exist in high-redshift galaxies. We also partially explore varying the gas metallicity.
Our goal is to run a realistic set of simulations of formation of star clusters in molecular clouds to understand the physics of star formation across cosmic time: from conditions typical of present-day ISM to the the higher-pressure environments found in the ISM of higher redshift galaxies. 

In this paper we focus on understanding the IMF, the SFR and SFE as a function of the cloud mass and compactness. We derive a star formation law valid at densities typical of high-redshift molecular clouds that will help to justify and inform the sub-grid star formation recipe used in cosmological simulations.

\setlength\extrarowheight{5pt}
\begin{table*}
  \caption{A collection of results. \label{tab:2}}
\begin{threeparttable}
\centering
  \def\arraystretch{1.1}
  \begin{tabular}{rccSccSScS}
    \toprule
    Cloud name  &
    $m_{\rm gas}(\msun)$ & 
    $\overline{n}_{\rm gas}(\si{cm^{-3}})$ &
    {$\Sigma \, (\msun ~ {\rm pc}^{-2})$} &
    $Z(Z_\odot)$ &
    $m_{cl}(\msun)$ \tnote{a} &
    {TSFE (\%)} \tnote{b} &
    {SFR$_{ff}$} \tnote{c} &
    IMF slope \tnote{d} &
    $n_{\rm SN}$ \tnote{e} 
    \\
    \midrule
    XS-F     & \num{3.2e+03} & \num{1.8e+02} & 41       & 1     & \num{3.8e+02} & 12.1  & 0.18 & $1.0^{+0.4}_{-0.3}$ &2     \\
S-F      & \num{1.0e+04} & \num{1.8e+02} & 61       & 1     & \num{5.1e+02} & 5.1   & 0.062 & $1.3^{+0.3}_{-0.3}$ &2     \\
M-F      & \num{3.2e+04} & \num{1.8e+02} & 89       & 1     & \num{1.4e+03} & 4.3   & 0.042 & $1.1^{+0.2}_{-0.2}$ &12    \\
L-F      & \num{1.0e+05} & \num{1.8e+02} & 131      & 1     & \num{5.7e+03} & 5.7   & 0.053 & $1.2^{+0.2}_{-0.1}$ &38    \\
XL-F     & \num{3.2e+05} & \num{1.8e+02} & 193      & 1     & \num{2.5e+04} & 7.8   & 0.043 & $1.1^{+0.1}_{-0.1}$ &142   \\
XS-C     & \num{3.2e+03} & \num{1.8e+03} & 193      & 1     & \num{1.0e+02} & 3.3   & 0.033 & $0.5^{+0.8}_{-0.0}$ &0     \\
S-C      & \num{1.0e+04} & \num{1.8e+03} & 283      & 1     & \num{5.3e+02} & 5.3   & 0.052 & $1.6^{+0.1}_{-0.3}$ &1     \\
M-C      & \num{3.2e+04} & \num{1.8e+03} & 415      & 1     & \num{3.0e+03} & 9.4   & 0.047 & $1.2^{+0.2}_{-0.2}$ &5     \\
L-C      & \num{1.0e+05} & \num{1.8e+03} & 609      & 1     & \num{1.4e+04} & 13.7  & 0.099 & $1.2^{+0.1}_{-0.1}$ &47    \\
L-C-lm   & \num{1.0e+05} & \num{1.8e+03} & 609      & 1/10  & \num{3.4e+03} & 3.4   & 0.021 & $1.2^{+0.2}_{-0.2}$ &5     \\
L-C-xlm  & \num{1.0e+05} & \num{1.8e+03} & 609      & 1/40  & \num{3.3e+03} & 3.3   & 0.025 & $1.0^{+0.2}_{-0.2}$ &5     \\
XXS-VC   & \num{1.0e+03} & \num{1.8e+04} & 609      & 1     & \num{9.8e+01} & 9.8   & 0.099 & $0.5^{+0.9}_{-0.0}$ &0     \\
XS-VC    & \num{3.2e+03} & \num{1.8e+04} & 894      & 1     & \num{5.1e+02} & 16.1  & 0.2 & $1.0^{+0.3}_{-0.2}$ &0     \\
S-VC     & \num{1.0e+04} & \num{1.8e+04} & 1312     & 1     & \num{3.2e+03} & 32.2  & 0.31 & $1.5^{+0.2}_{-0.2}$ &0     \\
M-VC     & \num{3.2e+04} & \num{1.8e+04} & 1925     & 1     & \num{1.5e+04} & 46.6  & 0.25 & $1.4^{+0.1}_{-0.1}$ &0     \\
L-VC     & \num{1.0e+05} & \num{1.8e+04} & 2827     & 1     & \num{2.7e+04} & 27.4  & & $1.3^{+0.1}_{-0.1}$ &0     \\

    \bottomrule
  \end{tabular}
\begin{tablenotes}
	\item[]
  (a) Stellar mass of the cluster formed from the cloud.
  (b) Total star formation efficiency, equal to $m_{cl}/m_{\rm gas}$.
  (c) Peak dimensionless star formation rate per free-fall time.
  (d) Negative IMF power-law slope $\Gamma$: $\md N/\md \log m \propto m^{-\Gamma}$.
  (e) Number of SNe explosions in 7 free-fall time of simulation.
\end{tablenotes}
\end{threeparttable}
\end{table*}
A summary of simulations results is presented in Table~\ref{tab:2}. The main findings of this paper are the following:
\begin{enumerate}
\item  We find that a Chabrier (or Krupa) stellar IMF with the correct normalization can can be reproduced in all of our simulations if we assume that each star-forming gas clump (sink particle) fragments into stars with a power-law mass function with log-slope $\Gamma \sim 0.8$, flatter than the mass function of the sink particles, which have Kroupa slope $\Gamma \sim 1.3$. With this prescription we find that statistically about $40 \%$ of the mass of the sink particle is locked into a single star, while the remaining $60 \%$ is distributed into smaller mass stars. This result is in agreement with the observed mass function of dense cores in some molecular clouds. \revi{The resolution study shows that increasing the resolution changes the CMF, but the total mass in cores remains nearly the same. For these reasons, we find that the model in which cores fragment with nearly 100\% efficiency into stars is the most likely model, although we cannot rule out alternative scenarios.}
\item  The IMF of stars at any time during the star formation burst is Chabrier-like. Because the total mass in stars is initially small and grows with time, at the beginning of the simulations, statistically, there are fewer high-mass stars. The apparent behaviour is that low and intermediate-mass stars form first, followed by the most massive stars.
\item  The star formation law that best describes star formation in molecular clouds found in the local universe (\ie, in fiducial simulations) is $d\rho_*/dt \approx 1.1\% \rho_{gas}/t_{ff}$. In dense molecular clouds with $\overline n_{\rm gas} >n_{cri} \approx 10^3$~cm$^{-3}$, more typically found in high-redshift galaxies, we find $d\rho_*/dt \approx 1.1\% \rho_{gas}^2/(\rho_{cri}t_{ff}) \propto \rho_{\rm gas}^{2.5}$. The duration of the star formation episode in all simulations is roughly 6 sound crossing times of the cloud radius (with $c_s = 10~\si{km/s}$).
\item For gas at solar metallicity the total star formation efficiency in the cloud is $f_{*, tot}=2\% (m_{\rm gas}/10^4~\msun)^{0.4}(1+\overline n_{\rm gas}/n_{\rm cri})^{0.91}$, where $n_{\rm cri} \approx 10^3$~cm$^{-3}$, also in agreement with {\it (iii)}. 
\item At metallicity $Z<0.1$~Z$_\odot$, $f_*$ is reduced by a factor of $\sim 5$ \revi{due to more efficient UV feedback caused by the higher temperature and pressure of \HII regions. We do not observe a dependence of the IMF on the metallicity, in agreement with previous studies.}
\item \revi{We note that the most compact and massive clouds appear to form globular cluster progenitors, in the sense that star clusters remain gravitationally bound after the gas has been mostly expelled. We plan to explore in detail the dynamics of these bound star clusters and possible relationships with the star formation efficiency and the escape fraction of ionising photons in future works.}
\end{enumerate}
The second paper of this series we will focus on
calculating the escape fraction of ionising photons, \fesc, from molecular clouds. This is the first necessary step for a realistic estimate of the escape fraction from galaxies.
Finally, in a third paper we will take a closer look at the dynamics of the star clusters and connect with important questions on the role of compact star clusters in creating seed black holes that might grow into supermassive black holes, and questions in Near Field Cosmology on the origin of globular clusters and ultra-faint dwarfs.

\section*{ACKNOWLEDGEMENTS}
We would like to thank the referee for the insightful comments that helped improve the quality of the paper.
MR acknowledges the support by NASA grant 80NSSC18K0527. 
The authors acknowledge the University of Maryland supercomputing resources (http://hpcc.umd.edu) made available for conducting the research reported in this paper.

This work has been funded by the European Research Council under the European Community's Seventh Framework Programme (FP7/2007-2013). SG has received funding from Grant Agreement no. 339177 (STARLIGHT) of this programme. 

\appendix

\section{Clump finder criteria}
\label{sec:app1}

\begin{figure}
  \centering
  \includegraphics[width=\columnwidth]{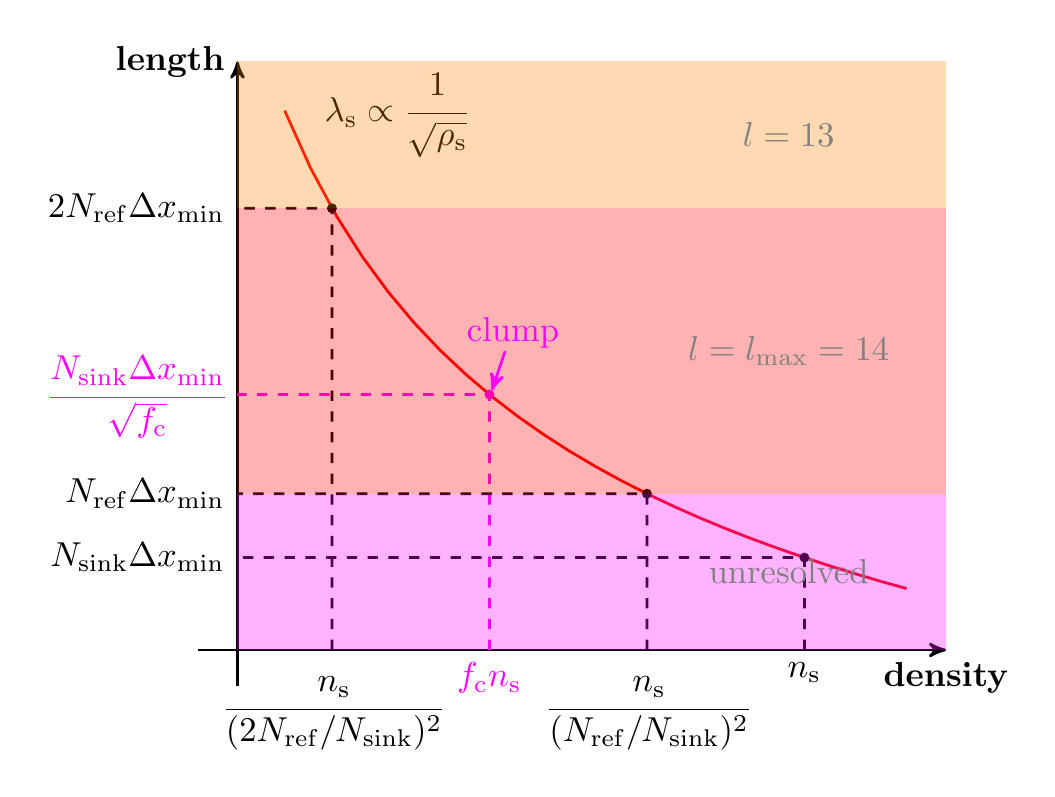}
  \caption{Explanation of the sink formation criteria in Equation~(\ref{eq:Nc}).
  The x-axis is the density of a given cell and the y-axis is the corresponding Jeans length.
  Refer to the text for the meaning of the labels. We impose that the clump finder acts at the highest refinement level but before the clump becomes unresolved.
   }
  \label{fig:refine}
\end{figure}
In this appendix we justify out choice for the value of $N_{\rm sink}=5$ in Section~\ref{ssec:sink}. We find that $N_{\rm sink}$ should be constrained by the relationship:
\begin{equation}
  \label{eq:Nc}
  N_{\rm ref} \sqrt{f_{\rm c}} < N_{\rm sink} < 2N_{\rm ref}\sqrt{f_{\rm c}},
\end{equation}
where $N_{\rm ref}$ is number of Jeans lengths for the refinement criteria, and $f_{\rm c}=1/10$ is the ratio of clump-finder threshold density to the sink threshold density. In our case, for $f_c=0.1$ and $N_{\rm ref}=10$, we have $3<N_{\rm sink}<6$. Therefore in all our simulations we set $N_{\rm sink}=5$ to satisfy Equation~(\ref{eq:Nc}).
The constraint in Equation~(\ref{eq:Nc}) can be understood by inspecting the sketch in Figure~\ref{fig:refine}, showing the Jeans length as a function of the gas density in a cell at different refinement levels (horizontal bands).  As the
gas density increases the Jeans length decreases and the level of
refinement increases up to the maximum level in the simulation (\eg, $n_{\rm refine}=14$). 
The clump finder has a lower density threshold than the sink formation threshold in order to identify structures that should form sinks. In order to ensure that these clumps are maximally resolved, we set all clumps to be at the highest refinement level. This gives the constraint
$ \frac{1}{(2N_{\rm ref}/N_{\rm sink})^2} < f_{\rm c} < \frac{1}{(N_{\rm
    ref}/N_{\rm sink})^2} $, and therefore Equation~(\ref{eq:Nc}) follows.

\section{Emission from clusters}
\label{sec:app2}
\begin{figure}
  \centering
  \includegraphics[width=\columnwidth]
  {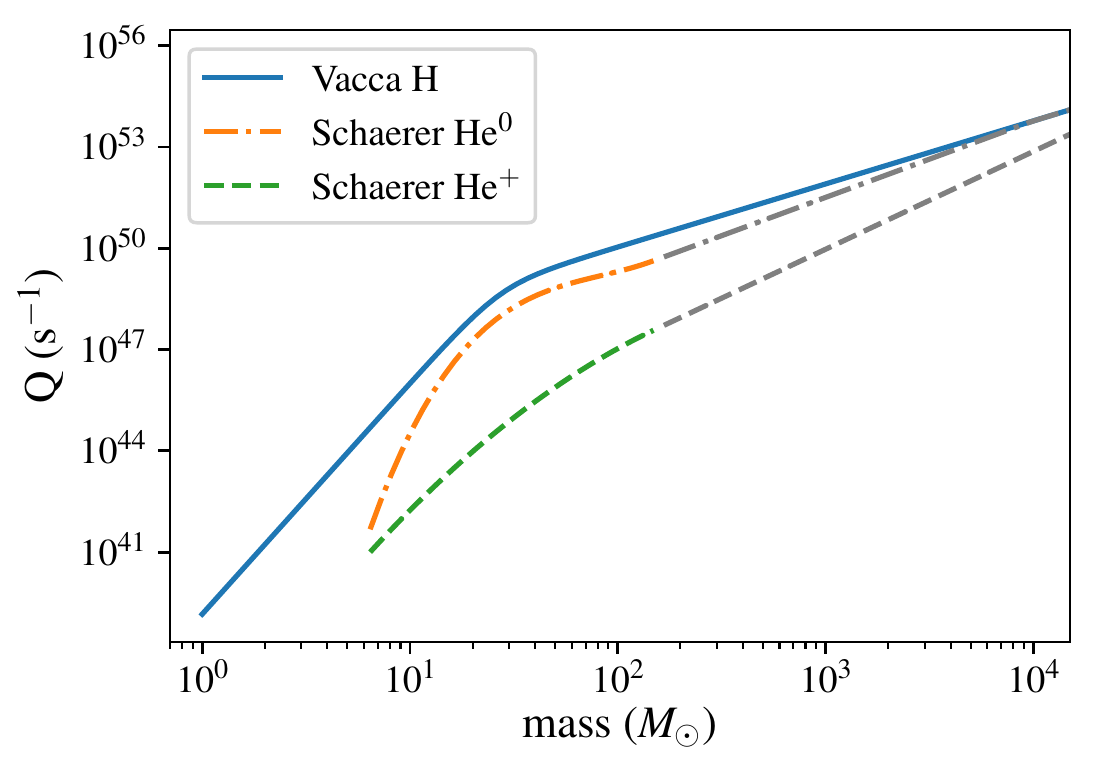}
  \caption{Ionising photon emission rate as a function of stellar mass.
  The colored lines are $Q_{\rm H}$ from Vacca fit
and $Q_{\rm He^0}$, $Q_{\rm He^{+}}$ from Schaerer fit.
The gray lines are their extrapolations.}
  \label{fig:Q}
\end{figure}

\begin{figure}
  \centering
  \includegraphics[width=\columnwidth]{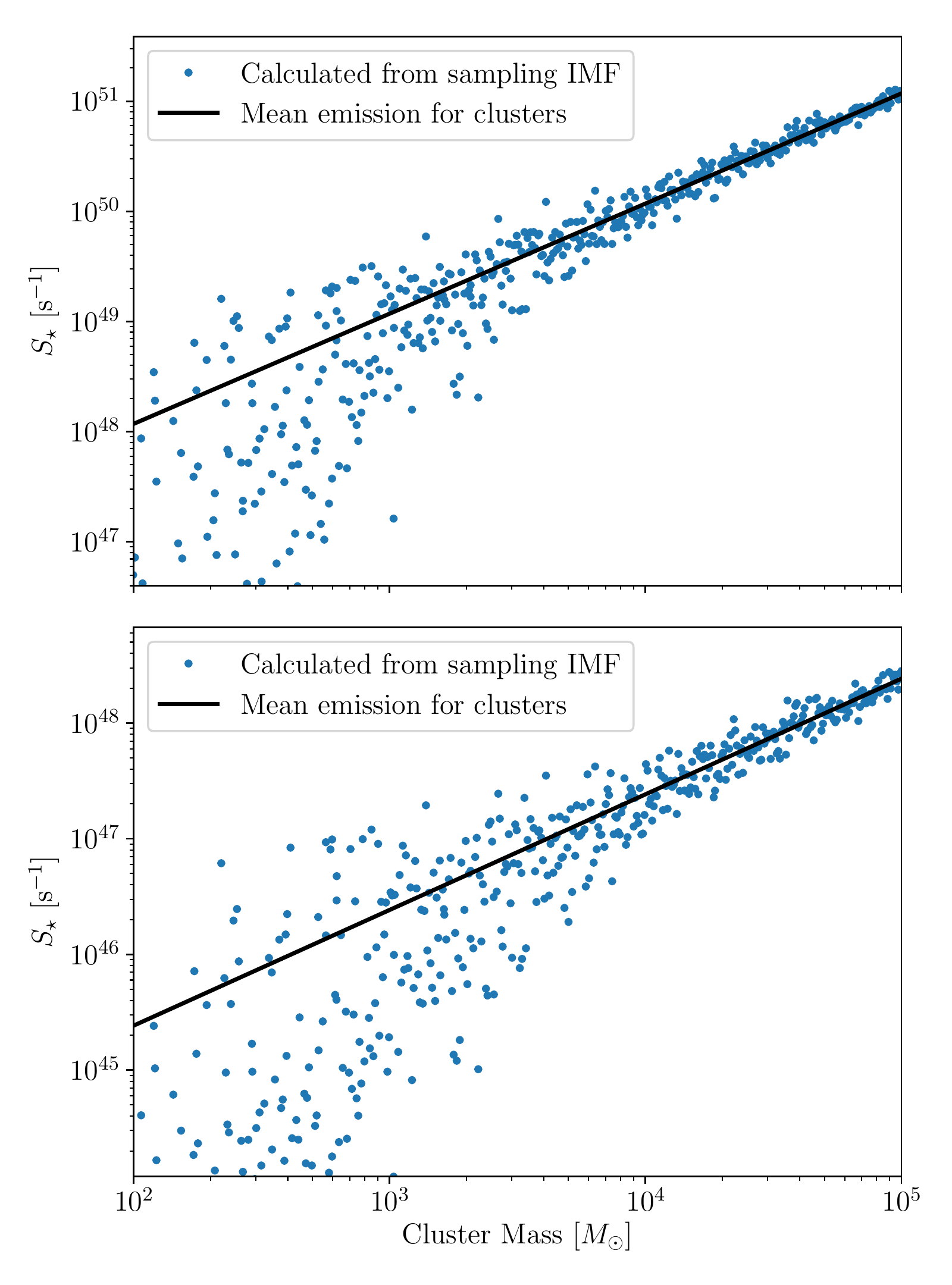}
  \caption{
  He$^0$ (top) and He$^+$(bottom) ionising photon emission rate as a function of the star cluster mass. The black solid lines are given by $S_* = k M_\ast$, where $M_\ast$ is mass of the star cluster and $k$ is $1.178 \times 10^{46}\;\mathrm{s}^{-1}M_\odot^{-1}$ and $2.422\times10^{43}\;\mathrm{s}^{-1}M_\odot^{-1}$ for He$^0$ and He$^+$, respectively.
  }
  \label{fig:Hefit}
\end{figure}

In this appendix, we estimate the approximate helium-ionising photon 
emission rate from stellar clusters of a range of masses.
The ionising photon emission rate from individual stars is plotted in Figure~\ref{fig:Q}.
We do a Monte Carlo sampling of clusters of stars with a Kroupa IMF and calculate the 
He$^0$ and He$^+$ ionising photon emission rates using \cite{Schaerer:2002} fit for each star. 
We assume a upper and lower limits of the star masses of $0.08M_\odot$ and $100M_\odot$.
These results are plotted in Figure~\ref{fig:Hefit}, along with a linear fit assuming a perfect sampling of the stellar population.

\bibliographystyle{mnras}
\bibliography{Bib_bibdesk.bib,samgeen.bib}

\bsp	
\label{lastpage}
\end{document}